\documentclass[pra, aps, 10pt, showkeys, twocolumn, nofootinbib, showpacs]{revtex4-1}
\usepackage[utf8]{inputenc}
\usepackage[english]{babel}
\usepackage{amsmath}
\usepackage{amssymb}
\usepackage{graphicx}
\usepackage{subfig}
\usepackage{hyperref}

\begin{document}

\title{Free energy analysis for a system of interacting particles arranged in Bravais lattice}

\author{B.\,I.~Lev}
\author{V.\,B.~Tymchyshyn}
\author{A.\,G.~Zagorodny}

\affiliation{Bogolyubov Institute for Theoretical Physics, National Academy of Science, Metrolohichna St. 14-b, Kyiv 03680, Ukraine}

\begin{abstract}
We propose a method of free energy calculation for a system of interacting particles arranged in a Bravais lattice.
It will be shown how to treat divergences for infinite unbounded systems with ``catastrophic'' potentials like Coulomb and compare two of them.
Besides, we show that current method may be used not only for essentially classical systems, but for some quantum as well.
Two systems are considered: grains in dusty plasma and electrons on the liquid-helium surface.
For dusty-plasma parameters of grain's lattice are calculated by numerical solution of obtained equations.
Electrons on the liquid-helium surface are analyzed to get localization distance.
Besides, conditions for existence of electron lattice are found.
\end{abstract}

\pacs{
52.27.Lw, 
71.10.Pm, 
02.30.Mv  
}

\maketitle

\section*{Introduction}

There are many soft-matter systems, such as grains in dusty plasma, colloids in various solvents, surfactant solutions, etc., that exhibit self-organization and rearrangement in crystalline structures under certain conditions.
Often their inter-particle potential is long-range and similar to the Coulomb one.
Usually, this kind of potentials cause divergences (that's why they are often called ``catastrophic'') and complicates consideration \cite{bilotsky,kantorovich_tupitsyn,buhler_crandal,borwein}.
On the other hand, mentioned systems are very interesting due to their application to the studies of a variety of peculiar phenomena in different fields of science \cite{fortov,lowen,morfill_tomas}.

It seems to be the most challenging problem to treat Coulomb-like systems with high concentrations of interacting particles \cite{helpand}.
When concentration increases, one can observe crystallization-like phenomena and transition between different lattice symmetries \cite{klumov_joyce,klumov_morfill,lev_zagorodny,lev_tymchyshyn_cond_matt,thomas_morfill}.

In current contribution we will treat concrete systems along with general formalism development.
They were chosen as manifesting a lot of interesting effects.
For example, dusty plasma may serve as perfect media for the experimental investigation of classical fluids and solids along with colloids \cite{thomas_morfill,chu_lin,melzer_trottenberg,vladimirov_khrapak,ikezi,melzer_homann,sitenko_zagorodny,brazovsky,lev_yokoyama,totsuji_kishimoto}.
The second system is ``electrons on the liquid-helium surface'' \cite{leiderer,shikin_leiderer,tsui_stormer,laughlin}.
It may undergo Wigner crystallization \cite{edelman,ando_fowler_stern,wigner,grimes_adams,platzman_fukuyama,skachko,lambert} and transition between triangular and square lattices \cite{haque,lev_tymchyshyn}.
Moreover, homogeneous distribution of electron density is not always stable, and there are critical parameters when spatial structures, especially periodic deformations and multi-electron dimples, are formed \cite{gorkov_chernikova}.

But our original goal is some general description of particles arranged in a Bravais lattice when interparticle potential may be ``catastrophic''.
It is rather difficult, since traditional methods of statistical mechanics cannot be applied to systems with Coulomb-like interactions because of free-energy divergences.
Moreover, partition function can be exactly evaluated only for few model systems of interacting particles in the thermodynamical limit \cite{ruelle,isihara,huang,baxter}.
Thus, basing on \cite{bilotsky_lev,lev_zhugayevich,lev_zagorodny_phys_rev,lev_tymchyshyn}, we will use alternative approach that employs Hubbard-Stratonovich representation of the partition function \cite{stratonovich}.
Besides, we use approach involving probability distribution function for the whole lattice as in \cite{bilotsky}, but construct it from single-particle distribution functions with a special procedure.
Namely, we will expand and significantly supplement results obtained in \cite{lev_tymchyshyn} for two-dimensional systems.
New results for three-dimensional case, that is far more complicated than two-dimensional, will be obtained.

To be more rigorous: our goal is to find expression for free energy of particles arranged in a Bravais lattice even if inter-particle potential is ``catastrophic''.
General equations will be applied to two concrete systems: three-dimensional\,--- grains in dusty plasma, and two-di\-men\-si\-o\-nal\,--- electrons on the liquid-helium surface.
In first case we'll minimize this energy to find out which lattice should be observed experimentally.
And for electrons localization distance will be found.
Besides, this will show that presented method allows easily reduce dimension number from three to two.

Present article is organized as follows.
It can be contingently divided into two parts: development of technique and its application.
Sections \ref{sec:stat_descr} and \ref{sec:lattice} focus on general expression for free energy for any Bravais lattice and any potential.
The former shows that not only classical, but lots of quantum systems may be treated this way.
Besides, an expression for ``entropy part'' of free energy is calculated.
And in section \ref{sec:lattice} interaction energy for paricles arranged in a Bravais lattice is calculated.

Both this sections are equipped with examples (\ref{subsec:two_dim_fermi}, \ref{subsec:pdf_for_sp}--\ref{subsec:capillary_interaction}) that show how to use developed formalism.
This examples are rather useful, because they are later reused when treating physical systems.

Concrete applications of developed formalizm start with section \ref{sec:dust_crystal}.
Here crystal of dust particles is considered.
In \ref{sec:dust_crystal} it is shown that hexagonal close packing seems to be the lattice we expect to be seen in experiment.
This is in agreement with experiments \cite{klumov_joyce} and computer simulations \cite{klumov_morfill}.

System of electrons on the liquid-helium surface is analyzed in \ref{sec:electrons_on_helium}.
In \ref{sec:electrons_on_helium} localization distance is calculated and compared to classical result \cite{edelman}.
For low temperatures $T \to 0$ they coincide.

For not to be confused with lots and lots of complicated mathematical computations we moved them to appendices \ref{sec:appendices}.
Thus article mostly contains some physical explanations and results.
But using references to appendices, one may find here rigorous exposition of presented ideas as well.

\section{Statistical description of interacting particles system}
\label{sec:stat_descr}

First of all let's outline class of systems we will be able to consider.
Since everything is clear with classical systems, we will give heed to quantum ones.

In this section we aim some general consideration of inhomogeneous system of interacting particles~\cite{lev_zhugayevich,lev_tymchyshyn}.
Nor system, neither interaction potential will be specified to keep generosity and extendability.
As result, we will get free energy in some general form and ready for ``applying'' translation symmetry (section~\ref{sec:lattice}).
Of course, along this section we will do some assumptions about system under consideration.
Together they will form restrictions on quantum systems we can treat with proposed method.

In current model macroscopic states of the system are described by occupation numbers.
Besides, we suppose that it's Hamiltonian has form:
\begin{equation}
\label{hamiltonian}
    H = \sum_s \varepsilon_s n_s + \frac{1}{2} \sum_{s,s'} V_{ss'} n_s n_{s'}.
\end{equation}
Here $\varepsilon_s$ is the additive part of particle energy (usually it is kinetic energy, but it can be particle's energy in external field as well.), $s$ indicates particle state, $V_{ss'}$ is interaction energy between particles in states $s$ and $s'$, $n_s$ is the occupation number of state $s$.
We neglect quantum correlations, so only essentially classical systems are considered.

Partition function for this kind of system will be:
$$
    Z = \sum_{\{n_s\}} \exp (-\beta H),
$$
where summation is performed over all possible states $\{n_s\}$ of the system.

Now we use some properties of Gaussian integrals over auxiliary fields, i.e. Hubbard-Stratonovich transformation \cite{stratonovich,hubbard}:
\begin{multline*}
\exp \left( 
         \frac{\nu^2}{2 \vartheta} \sum_{s,s'} \omega_{ss'} n_s n_{s'} 
     \right) =\\=
\int\limits_{-\infty}^{\infty} D\varphi 
\exp \Biggl( 
    \nu \sum_s n_s \varphi_s
  - \frac{\vartheta}{2} \sum_{s,s'} \omega_{ss'}^{-1} \varphi_s \varphi_{s'} 
     \Biggr),
\end{multline*}
with 
$D \varphi = \prod_s d\varphi_s/\sqrt{\det(2 \pi \beta \omega_{ss'})}$. 
Second-order dependence on occupation numbers is avoided by carrying them to introduced field and partition function can be written:
$$
Z = \!\!\!\int\limits_{-\infty}^{\infty}\!\!\!
      D\varphi 
      \exp\!\! \left[
          \sum_s (i \varphi_s - \beta \varepsilon_s) n_s 
        - \frac{1}{2\beta} \sum_{s,s'} \left(V_{ss'}^{-1} \varphi_s \varphi_{s'} \right) 
      \right]\!.
$$
For canonical ensemble number of particles can be fixed by using Cauchy equation:
$$
    \frac{1}{2 \pi i} \oint \xi^{\sum_s n_s - N - 1} d\xi = 1,
$$
that leads to $N$-particle partition function:
$$
\begin{aligned}
Z_N &= \frac{1}{2 \pi i} 
       \oint d\xi \int\! D\varphi \exp\Biggl[ 
          - \frac1{2\beta} \sum_{s,s'}
                V_{ss'}^{-1} \varphi_s \varphi_{s'} - 
       \Biggr.\\
    &  \Biggl. 
          - (N+1) \ln\xi 
       \Biggr] \prod_s \sum_{\{n_s\}} \left[
           \xi \exp(i \varphi_s - \beta \varepsilon_s)
       \right]^{n_s}.
\end{aligned}
$$

Performing summation over occupation numbers:
\begin{equation}
\label{betaF0}
\begin{aligned}
Z_N &= \frac{1}{2 \pi i} \oint d\xi \int D\varphi \exp [- \beta F(\varphi,\xi)],\\
  \beta F&(\varphi,\xi) 
= \frac1{2} \sum_{s,s'} V_{ss'}^{-1} \varphi_s \varphi_{s'} + \\
&+ \delta \sum_s \ln \left(
       1 - \delta \xi e^{-\beta \varepsilon_s+i \varphi_s}
   \right) + (N+1) \ln \xi . 
\end{aligned}
\end{equation}
Here type of statistics is incorporated into $\delta$\,--- it equals to $+1$ for Bose-Einstein, $0$ for Maxwell-Boltzmann and $-1$ for Fermi-Dirac statistics.

Now we have expression for free energy in auxiliary field representation with $\xi = \exp(\beta\mu)$ for chemical activity.
Equation~\eqref{betaF0} contains all information about probable states of the system and corresponds to the sequence of equilibrium states with regard to their weights.
We are interested in asymptotic value of partition function, but want to avoid using perturbation theory.
Thus domain is extended to complex plane and saddle-point method is applied.

Dominant contribution is made by states satisfying extrema condition:
$$
  \frac{\delta \beta F}{\delta \varphi}
= \frac{\delta \beta F}{\delta \xi} 
= 0.
$$
Variation of~\eqref{betaF0} yields expression for saddle-point states:
\begin{subequations}
\begin{align}
  &\frac{1}{\beta} \sum_{s'} V_{ss'}^{-1} \varphi_{s'} 
- \cfrac{i \xi e^{-\beta\varepsilon_s + i \varphi_s}}
       {1 - \delta\xi e^{-\beta\varepsilon_s + i \varphi_s}} 
= 0; \label{saddle1}\\
  &\sum_s \cfrac{\xi e^{-\beta \varepsilon_s + i \varphi_s}}
              {1 - \delta \xi e^{-\beta \varepsilon_s + i \varphi_s}} 
= N+1.  \label{saddle3}
\end{align}
\end{subequations}
For a certain state expression
\begin{equation}
  \label{f_s}
  f_s = \frac{\xi e^{-\beta\varepsilon_s + i \varphi_s }}
           {1 - \delta\xi e^{-\beta\varepsilon_s + i \varphi_s}},
\end{equation}
from \eqref{saddle3} can be treated as an average occupation number.
Now saddle-point states (that may be interpreted as thermodynamically stable distributions) can be obtained.

Equation~\eqref{saddle1} contains inverse matrix, that is quite inconvenient for us, because its calculation is a challenging mathematical problem even for simple potentials~\cite{lev_zhugayevich,edwards_lennard,samuel}.
This problem can be avoided if we use~\eqref{f_s} and perform inverse transformation
$$
\begin{aligned}
&\varphi_s = i \beta \sum_{s'} V_{ss'} f_{s'},\\
&\frac1{2\beta} \sum_{s,s'} V_{ss'}^{-1} \varphi_s \varphi_{s'} 
  = - \frac{\beta}2 \sum_{s,s'} V_{ss'} f_s f_{s'}.
\end{aligned}
$$
Than free energy can be written
\begin{equation}
\label{betaF1}
\begin{aligned}
 \beta F[f,\xi] 
= - \frac{\beta}{2} \sum_{s,s'} V_{ss'} f_s f_{s'} 
 &- \delta \sum_{s} \ln (1 + \delta f_s) \\
 &+ (N+1) \ln \xi (f).
\end{aligned}
\end{equation}
In terms of canonical ensemble we can write from~\eqref{f_s}
$$
\begin{aligned}
      \ln\xi(f)
   &= \frac{1}{N}\sum_s f_s
      \left[ 
          \beta(\varepsilon_s + E_s) + \ln f_s - \ln (1 + \delta f_s) 
      \right],\\
      E_s
   &= \sum_{s'} V_{ss'} f_{s'},
\end{aligned}      
$$
and substitute last expression into free energy~\eqref{betaF1} to rewrite its expression without chemical potential.
\begin{equation}
\label{free_energy}
\begin{aligned}
   F[f] 
&= \sum_s f_s \varepsilon_s 
 + \frac{1}{2} \sum_{s,s'} V_{ss'} f_s f_{s'} + \\
&+ \underbrace{\frac{1}{\beta} \sum_s \left[f_s \ln f_s - (f_s+\delta) \ln (1 + \delta f_s) \right]}_{F_{ent}}.
\end{aligned}
\end{equation}

First two terms in~\eqref{free_energy} are kinetic and potential energy respectively, the last one is a contribution due to entropy $F_{ent}$ (should be equal to zero if $T=0$).

Now let's do a quick test of obtained results.
If a grand canonical ensemble with fixed chemical potential is considered, we can get from~\eqref{f_s}
\begin{equation}
    \label{rozp_uzag}
    f_s = \frac{1}{e^{\beta (\varepsilon_s - \mu_s) } - \delta},
\end{equation}
which is generalization of the well-known distribution with a chemical potential
$$
    \mu_s = \mu - E_s.
$$
It is obvious that the ``saddle point'' and ``mean field'' approximations are equivalents in this case.
If an ideal gas ($\mu_s \equiv \mu$) is considered, we can obtain classical statistical distributions.

Before winding up this section let's review the results.
We aimed to find some general expression for free energy in a mean-field approximation.
As the one, equation~\eqref{free_energy} should be pointed out (and in some sense~\eqref{rozp_uzag} that can be used to obtain particles distribution in grand canonical ensemble).
Besides, we have got some restrictions on the class of systems we will be able to consider: special form of Hamiltonian~\eqref{hamiltonian} and negligible quantum cross-correlations.

\subsection{Example: two-dimensional system of localized Fermi particles\texorpdfstring{\,---}{-} electrons on the liquid-helium surface}
\label{subsec:two_dim_fermi}

Before we shift our attention to potential energy of particles arranged in a lattice, let's do some example.
With this subsection we aim two things: give some idea on how to use the above formalism and secondly\,--- calculate contribution due to entropy for electrons on the liquid-helium surface (will be used in~\ref{sec:electrons_on_helium}).

Electrons on the liquid-helium surface have two degrees of freedom only~\cite{edelman,ando_fowler_stern}.
Though we set $\delta = -1$ (Fermi particles) and $d=2$ (system is two-dimensional).
Then equation~\eqref{free_energy} can be written
\begin{equation}
\label{f_mu_nonintegrated}
  \begin{aligned}
    &\beta F[\mu] = \int \frac{d^2 \vec{p}\, d^2 \vec{r}}{(2 \pi \hbar)^2} \frac{\beta \vec{p}^{\,2}}{2 m} 
                         \frac{1}{e^{\beta(\vec{p}^{\,2} / (2 m) - \mu(\vec{r}\,))} + 1} + \\
    & {} + \frac{\beta}{2} \iint \frac{d^2 \vec{p}\, d^2 \vec{r}}{(2 \pi \hbar)^2} 
           \frac{d^2 \vec{p}^{\,\prime}\, d^2 \vec{r}^{\,\prime}}{(2 \pi \hbar)^2} \times \\
    & {} \times \frac{V\left(|\vec{r}-\vec{r}^{\,\prime}|\right)}
                     {
                          \left(e^{\beta(\vec{p}^{\,2} / (2 m) - \mu(\vec{r}\,))} + 1\right)\!\!
                          \left(e^{\beta(\vec{p}^{\,\prime 2} / (2 m) - \mu(\vec{r}^{\,\prime}\,))} + 1\right)
                     } -\\ 
    & {} - \int \frac{d^2 \vec{p}\, d^2 \vec{r}}{(2 \pi \hbar)^2} 
           \left( 
               \frac{\ln\left(e^{\beta(\vec{p}^{\,2} / (2 m) - \mu(\vec{r}\,))} + 1\right)}
                    {e^{\beta(\vec{p}^{\,2} / (2 m) - \mu(\vec{r}))}+1} \right. + \\
    &{} + \left.
               \frac{\ln\left(e^{-\beta(\vec{p}^{\,2} / (2 m) - \mu(\vec{r}\,))} + 1\right)}
                    {e^{-\beta(\vec{p}^{\,2} / (2 m) - \mu(\vec{r}\,))} + 1} 
          \right).
  \end{aligned}
\end{equation}
Here integration sign $\int$ means integration over the whole phase space.

Since last expression is quite cumbersome we should simplify it somehow.
As first approximation we suppose
$$
  \varepsilon_s = \varepsilon(p) = p^2 / (2m).
$$
Actually, in the presence of an external field, dispersion relation should be a bit different.
But we know, that system of Fermi particles, we will be interested in, is highly degenerated~\cite{edelman}.
So quadratic form for dispersion relation looks quite reasonable. 

Using last equation and introducing thermal length
\begin{equation}
\label{lambda_t}
  \lambda_T = \sqrt{2 \pi^2 \hbar^2 \beta / m},
\end{equation}
we get from~\eqref{rozp_uzag} following expression
\begin{equation}
\label{rho_to_mu}
\begin{aligned}
  \rho(\vec{r}\,) &= \int \frac{d^2 \vec{p}}{(2 \pi \hbar)^2} \frac{1}{e^{\beta (\vec{p}^{\,2}/(2m) - \mu(\vec{r}\,))}+1} \\
                  &= \frac{\pi}{\lambda_T^2} \ln \left(1 + e^{\beta \mu(\vec{r}\,)}\right),
\end{aligned}
\end{equation}
which connects chemical potential $\mu(\vec{r}\,)$ and particles density $\rho(\vec{r}\,)$.

Now we return to simplification of~\eqref{f_mu_nonintegrated}.
Performing integration of~\eqref{f_mu_nonintegrated} over momentum and taking \eqref{rho_to_mu} and \eqref{lambda_t} into account one may get (see appendix~\ref{subsec:appendix_momentum_integration})
\begin{equation}
\label{free_energy_final}
  \begin{aligned}
     F[\rho] 
    &= \frac{1}{2} \iint d^2 \vec{r}\, d^2 \vec{r}^{\,\prime} V(|\vec{r}-\vec{r}^{\,\prime}|)
     \rho(\vec{r}\,) \rho(\vec{r}^{\,\prime}) + \\ 
    &+ \frac{\pi}{\beta\lambda_T^2} \int d^2 \vec{r}
     \left[ 
             \text{Li}_2\left(e^{- \pi^{-1} \lambda_T^2 \rho(\vec{r}\,)}\right) - \frac{\pi^2}{6}
     \right] +\\
    &+ \frac{\lambda_T^2}{2\pi\beta} \int d^2 \vec{r}
     \rho^2(\vec{r}\,).
  \end{aligned}
\end{equation}

One may notice that in case of Bose statistics we will loose the Bose-condensation effects due to this integration.
But we aim to use this equations for Fermi particles, so there is nothing to be worried about.

We know that electrons on the liquid-helium surface exist in forms of fluid or Wigner crystal~\cite{wigner,grimes_adams,platzman_fukuyama}.
We are interested in case when electrons are strongly localized, i.e. their coordinates are quite determined.
But since temperature differs from zero, we expect that particle's position will fluctuate near its equilibrium location.
Though we assume following distribution function $\rho_{\text{sp}}$ for \textit{\textbf{s}ingle \textbf{p}article} (somewhat analogous to form factor in QFT)
\begin{equation}
\label{rho_sp}
    \rho_{\text{sp}}(\vec{r}\,) = \frac{1}{(2\pi s^2)^{d/2}} e^{-r^2/\left(2s^2\right)},
\end{equation}
where $s$ is dispersion, or localization distance by physical meaning, and $d=2$ because system is two-dimensional\footnote{
We aim to reuse this and some of subsequent equations for systems with different number of dimensions. 
In such cases we will write them as functions of $d$ (dimensionality) even if it is redundant at the point we introduce the equation first time.}.

Equation~\eqref{rho_sp} is normal distribution which seems to be quite reasonable.
At least this distribution is valid for ground state of quantum harmonic oscillator which is the simplest approximation for particle fluctuating near its equilibrium location.

Last assumption we do is that ``Gaussian'' \eqref{rho_sp} is very ``sharp'' and every particle lays outside localization radius of any other particle.
This means that $s$ is much smaller then other characteristic distances in this system.
So free energy per one particle can be written:
\begin{equation}
\label{entropy_part}
  \begin{aligned}
       F_{\text{sp}}
    &= \frac{1}{2} \iint d^2 \vec{r}\, d^2 \vec{r}^{\,\prime} V(|\vec{r}-\vec{r}^{\,\prime}|)
       \rho_{\text{sp}}(\vec{r}\,) \rho(\vec{r}^{\,\prime}) + \\
    &+ \underbrace{\frac{\lambda_T^2}{8\pi^2 s^2 \beta} - \frac{\pi^4 s^2}{6\beta\lambda_T^2}.}_{F_{ent}}
  \end{aligned}
\end{equation}

Here $F_{ent}$ is contribution to free energy by \textit{\textbf{ent}ropy} and temperature.
First summand in $F_{ent}$ is obtained from direct integration of $\rho_{\text{sp}}^2$ (see last term in~\eqref{free_energy_final}).
To get second summand (penultimate term in~\eqref{free_energy_final}) we use ``sharpness'' of~\eqref{rho_sp} and approximate
\begin{equation}
 \text{Li}_2 \left(e^{- \pi^{-1} \lambda_T^2 \rho_{\text{sp}}(\vec{r}\,)}\right) \approx
 \left\{
 \begin{aligned}
    0       &,~ r \leq s;\\
    \pi^2/6 &,~ r > s.
 \end{aligned}
 \right.
\end{equation}

With equations~\eqref{entropy_part} and~\eqref{free_energy_final} we finish our consideration.
We fulfilled our goal to show an example of using section's~\ref{sec:stat_descr} formalism.
Moreover, now we have expression for entropy part $F_{ent}$ of free energy \eqref{entropy_part}, which we use later in~\ref{sec:electrons_on_helium}.

\section{Particles arranged in a lattice}
\label{sec:lattice}

In section~\ref{sec:stat_descr} we found some expressions for free energy of inhomogeneous system of interacting particles~\eqref{free_energy}.
In this section we will be interested in second term of~\eqref{free_energy}, i.e. potential energy.
More accurately, we aim to find method of potential energy calculation for system of identical (from inter-particle's potential point of view) particles arranged in a Bravais lattice.
During this section we will show how to use translation symmetry to get in some cases more convenient expressions for potential energy.

Last remark before we dive in.
We aim to find appropriate expressions for both two- and three-dimensional cases.
But to avoid reduplication all along the section~\ref{sec:lattice} only three-dimensional case will be treated as far more complicated.
For two-dimensional case we will only provide results and short description where difference comes from.

\begin{figure}[ht]
\centering
\includegraphics[width=8cm]{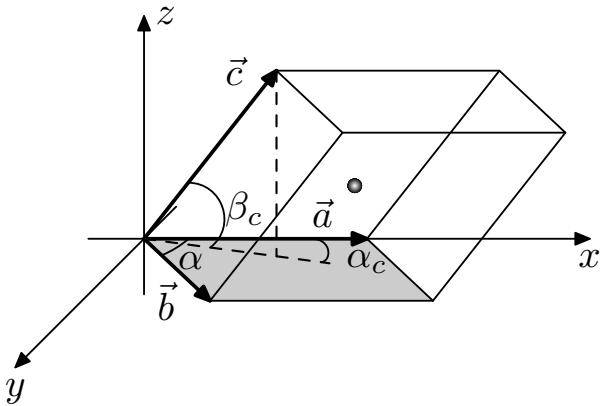}
\caption{
One cell from exact cover $\mathbb{V}$.
Vectors $\vec{a}$, $\vec{b}$ and $\vec{c}$ are base vectors for Bravais lattice.
Angle between $\vec{a}$ and $\vec{b}$ is supposed to be $\alpha$.
Angle between $\vec{c}$ and $XY$ plane is $\beta_c$ and its projection on $XY$ plane and $\vec{a}$ is $\alpha_c$.
One particle is shown in the center.
}\label{fig:vectors}
\end{figure}

Let $\mathbb{V}$ designates one subset of exact cover of $\mathbb{R}^3$ (figure~\ref{fig:vectors}) with similar domains containing strictly one particle (relative position of the particle supposed to be the same in every domain)\footnote{
For example, Wigner-Seitz cells provide this kind of covering.
But they often have too complicated shape and we will use parallelepipeds instead (Figure~\ref{fig:vectors}).}.
Then free energy of the particle localized near Bravais lattice site can be written as (we neglect kinetic energy in~\eqref{free_energy} since particles are localized and potential energy should be bigger)
\begin{subequations}
\label{freeEnergySeparated}
\allowdisplaybreaks
\begin{eqnarray}
   F_{int}       &=& \iiint\limits_{\mathbb{V}} \iiint\limits_{\mathbb{R}^3}
                         V\left(\left|\vec{r}-\vec{r}^{\,\prime}\right|\right)
                         \rho\left(\vec{r}\,\right) \rho\left(\vec{r}^{\,\prime}\right) 
                     d^3 \vec{r}^{\,\prime}\, d^3 \vec{r},\label{freeEnergySeparated-int}\\
   F_{s}         &=& \iiint\limits_{\mathbb{V}} \iiint\limits_{\mathbb{V}}
                         V\left(\left|\vec{r}-\vec{r}^{\,\prime}\right|\right)
                         \rho\left(\vec{r}\,\right) \rho\left(\vec{r}^{\,\prime}\right) 
                     d^3 \vec{r}^{\,\prime}\, d^3 \vec{r},~~~~~~\label{freeEnergySeparated-self}\\
   F_{\text{sp}} &=& F_{int} - F_{s} + F_{ent}. \label{freeEnergySeparated-sum}
\end{eqnarray}
\end{subequations}

Equation~\eqref{freeEnergySeparated} needs some explanations.
Interaction energy is presented here with $F_{int} - F_{s}$.
This ``splitting'' is just another way of saying we want to integrate over $\mathbb{R}^3 \setminus \mathbb{V}$.
If one collects both expressions together and uses $\iiint_{\mathbb{R}^3 \setminus \mathbb{V}} = \iiint_{\mathbb{R}^3} - \iiint_{\mathbb{V}}$ he will get regular integration over $\mathbb{R}^3 \setminus \mathbb{V}$ for $\vec{r}^{\,\prime}$.

In other words, since particle is localized near lattice site, we take some ``part of space'' $\mathbb{V}$ around it.
Then by integration we calculate interaction energy between particle in $\mathbb{V}$ and all the rest particles in $\mathbb{R}^3 \setminus \mathbb{V}$.

One may get a good intuitive idea about integration domain $\mathbb{R}^3 \setminus \mathbb{V}$ through discrete case.
When we calculate Coulomb energy for $i$-th particle we write $\sum_{j \neq i} e_ie_j/r_{ij}$.
Integration over $\mathbb{R}^3 \setminus \mathbb{V}$ is somewhat analogous to $j \neq i$ for discrete case.

We will further call $F_{int}$~\eqref{freeEnergySeparated-int} \textit{\textbf{int}eraction energy} and $F_s$~\eqref{freeEnergySeparated-self} \textit{\textbf{s}elf-interaction energy}.
This ``splitting'' has one more convenience.
Basing on physical background of considered potential we may want or not want to compensate self-interaction.
Obviously, Coulomb potential should have self-interaction compensated\,--- potential energy is zero if there is only one particle in a system (considered later in example~\ref{subsec:coulomb_potential}).
On the other hand, self-interaction for effective potentials caused by influence through medium should be uncompensated, because even in absence of other particles medium will influence this one.
Surface distortion is a good example.
If particle has distorted a surface it has already done some contribution to system's free energy independently on other particles (considered later in example~\ref{subsec:capillary_interaction}).
Of course, it interacts with distortions created by other particles.
But at the same time it interacts with ``its own'' surface distortion and this may be treated as ``self-interaction''.

Now lets consider~\eqref{freeEnergySeparated} more carefully.
All 3D Bravais lattices can be treated as constructed from parallelepipeds.
We will use this property and use parallelepiped as integration region $\mathbb{V}$.

Next thing we want to do is expressing probability distribution function $\rho$ for the whole $\mathbb{R}^3$ through single-particle probability distribution $\rho_{\text{sp}}$.
Basing on the inverse lattice vectors (appendix \ref{subsec:appendix_properties_of_f_k}) we will provide following decomposition
\begin{subequations}
\label{PDF_in_Fourier}
\allowdisplaybreaks
\begin{align}
      \rho(\vec{r}\,)              &= \sum\limits_{\vec{k}\in\mathbb{Z}^3} \rho_{\vec{k}} \mathfrak{f}_{\vec{k}}(\vec{r}\,), \label{PDF_in_Fourier_rho_xy}\\
      \rho_{\vec{k}}               &= \bar{\rho} \iiint\limits_{\mathbb{V}} \mathfrak{f}_{\vec{k}}^{*}(\vec{r}\,) \rho_{\text{sp}}(\vec{r}\,) d^3\vec{r}, \label{PDF_in_Fourier_rho_k}\\
 \mathfrak{f}_{\vec{k}}(\vec{r}\,) &= e^{2\pi\mathbf{i}\left(\vec{k}^{\,T}\hat{G}\vec{r}\,\right)}, \label{PDF_in_Fourier_f_k}\\
      \hat{G}                      &= \begin{pmatrix}
                                           \cfrac{1}{a} & -\cfrac{\cot(\alpha)}{a} & -\cfrac{\cot(\beta_c)\sin(\alpha-\alpha_c)}{a\sin(\alpha)}\\[12pt]
                                           0            & \cfrac{\csc(\alpha)}{b}  & -\cfrac{\cot(\beta_c)\sin(\alpha_c)}{b\sin(\alpha)}       \\[12pt]
                                           0            & 0                        & \cfrac{\csc(\beta_c)}{c}
                                      \end{pmatrix}. \label{PDF_in_Fourier_G_operator}
\end{align}
\end{subequations}
Here $*$ designates complex conjugation.
It may be noticed that all vectors are treated as column vectors.
$\bar{\rho}$ is mean particle density here.
To get better intuitive understanding how to compute~\eqref{PDF_in_Fourier} one may check example~\ref{subsec:pdf_for_sp} for $\rho_{\text{sp}}$ from \eqref{rho_sp}.

\begin{figure}[ht]
\centering
\includegraphics[width=2cm]{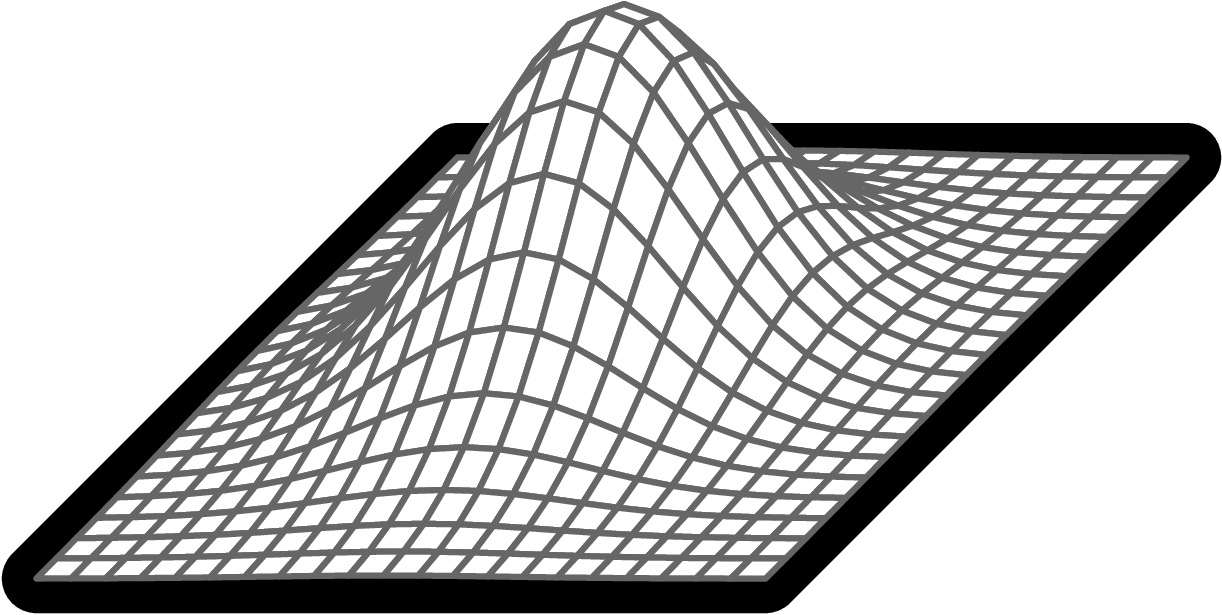}
\includegraphics[width=6cm]{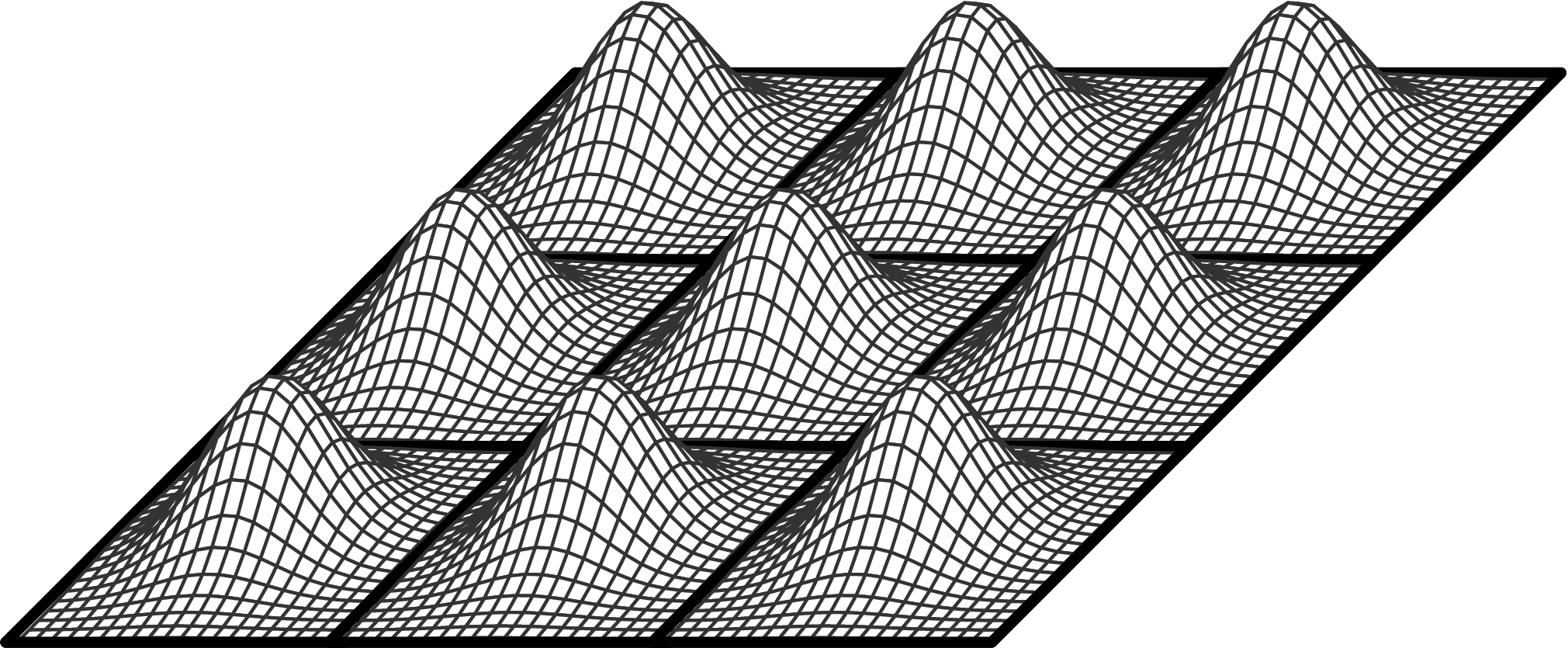}
\caption{
Two-dimensional analogy of transition from single-particle density probability function $\rho_{\text{sp}}(\vec{r}\,)$ to many particle $\rho(\vec{r}\,)$.
Function $\rho(\vec{r}\,)$ is defined for all $\mathbb{R}^2$ with two-dimensional analogy of \eqref{PDF_in_Fourier}.
}\label{fig:one_to_many}
\end{figure}

It may be shown that presented series are actually Fourier series for $\rho_{\text{sp}}$ and $\rho$ is periodic along vectors $\vec{a}$, $\vec{b}$ and $\vec{c}$ (appendix~\ref{subsec:appendix_properties_of_f_k}).
So we may treat $\rho\left(\vec{r}\,\right)$ as function on $\mathbb{R}^3$ ``composed'' from single particle distributions ``arranged in a lattice''.
This principle is demonstrated by Fig.~\ref{fig:one_to_many}.
Since $\rho$ has the same symmetry as the lattice and locally is a good approximation for $\rho_{\text{sp}}$ it will be treated as probability distribution function for the whole lattice.

Matrix $\hat{G}$ comprises all \textit{\textbf{g}eometry} of the lattice under consideration.
Performing minimization of $F_{\text{sp}}$ over entries of $\hat{G}$, one may find which lattice can be formed under certain conditions.
We will use later this type of analysis for dust crystal in section~\ref{sec:dust_crystal}.

If we use equations \eqref{PDF_in_Fourier} we can express \eqref{freeEnergySeparated-int} in terms of $\rho_{\text{sp}}$ (appendix~\ref{subsec:appendix_expression_f_int})
\begin{subequations}
\label{f_int}
\allowdisplaybreaks
\begin{align}
 F_{int}      &= \frac{1}{\bar{\rho}^{\,2}} \sum\limits_{\vec{k}\in\mathbb{Z}^3} |\rho_{\vec{k}}|^2\, V_{\vec{k}},\label{f_int_prime}\\
 V_{\vec{k}}  &= \bar{\rho} \iiint\limits_{\mathbb{R}^3}
                                \mathfrak{f}_{\vec{k}}\left(\vec{r}^{\,\prime}\right)
                                V\left(\left|\vec{r}^{\,\prime}\right|\right)
                            d^3 \vec{r}^{\,\prime}, \label{f_int_v_k}
\end{align}
\end{subequations}
where $\bar{\rho}$ is mean particle density.

Before moving further, one more thing should be emphasized.
This equations are useful because they allow us to deal with ``catastrophic'' potentials like Coulomb.
One may notice that for any descending potential only
$$
    \frac{1}{\bar{\rho}^{\,2}} |\rho_{0}|^2\, V_{0} = 4\pi \bar{\rho} \int\limits_0^\infty V(r) r^2 dr,
$$
can be diverging and only if potential is ``catastrophic''.

But this term does not contain ``lattice geometry'', it depends on mean particle density only.
This means, we can compare two lattices with equal mean particle density even if inter-particle potential is ``catastrophic'' (somewhat analogous to renormalization in QFT).

With equations~\eqref{f_int}, \eqref{PDF_in_Fourier} and~\eqref{freeEnergySeparated} we achieve our goal and finish consideration of interaction energy for particles arranged in a lattice.
Following examples show how to use obtained equations for screened Coulomb (example~\ref{subsec:coulomb_potential}) and capillary interaction (example~\ref{subsec:capillary_interaction}).
They both use $\rho$ obtained with $\rho_{\text{sp}}$ from~\eqref{rho_sp} (example~\ref{subsec:pdf_for_sp}).

\subsection{Example: calculating distribution function \texorpdfstring{$\rho$}{} for Gaussian single-particle distribution \texorpdfstring{$\rho_{\text{sp}}$}{}}
\label{subsec:pdf_for_sp}

We suppose that all particles are arranged in a lattice and their positions are quite determined.
But since temperature differs from zero we can expect that particle position fluctuates near its lattice site.
Using considerations from~\ref{subsec:two_dim_fermi} we suppose it to be~\eqref{rho_sp} with $d=3$.

Introduced constant $s$ means dispersion of presented distribution.
From physical point of view it can be treated as localization distance.
Later, in section~\ref{sec:electrons_on_helium}, we'll find its value and compare it with localization distance obtained in different works.

Using that $\rho_{\text{sp}}(r)$ is a very ``sharp'' Gaussian, we expand integration limits to $\mathbb{R}^3$ in~\eqref{PDF_in_Fourier_rho_k} and perform integration (appendix~\ref{subsec:appendix_rho_k})
\begin{equation}
\label{rho_k}
 \rho_{\vec{k}} = \bar{\rho} e^{-2 \pi^2 s^2 \vec{k}^{\,T} \hat{G} \hat{G}^{T} \vec{k}}.
\end{equation}

As mentioned previously, we are interested in two-dimensional case as well.
One may check, that there exists some sort of analogy of~\eqref{PDF_in_Fourier} for two dimensions \cite{lev_tymchyshyn}.
Difference will appeare in changing summations and integrations $\sum_{\mathbb{Z}^3} \rightarrow \sum_{\mathbb{Z}^2}$, $\iiint_{\mathbb{R}^3} \rightarrow \iint_{\mathbb{R}^2}$ and usage of different matrix
\begin{equation}
\label{two_dimensional_matrix}
      \hat{\mathcal{G}} = \begin{pmatrix}
                          1/a & -\cot(\alpha)/a \\[12pt]
                          0   & \csc(\alpha)/b  \\[12pt]
                          \end{pmatrix}.
\end{equation}

Performing calculations as in appendix~\ref{subsec:appendix_rho_k} with~\eqref{rho_sp} and $d=2$, one will get
\begin{equation}
\label{two_dimensional_rho_k}
 \rho_{\vec{k}} = \bar{\rho} e^{-2 \pi^2 s^2 \vec{k}^{\,T} \hat{\mathcal{G}} \hat{\mathcal{G}}^{T} \vec{k}},
\end{equation}
where $\bar{\rho}$ is two-, not three-dimensional density as in~\eqref{rho_k}.

Equations~\eqref{rho_k} and~\eqref{two_dimensional_rho_k} are aimed result of this example.
Later they will be used in calculations of potential energy for Coulomb~\eqref{subsec:coulomb_potential} and capillary~\eqref{subsec:capillary_interaction} interactions.

\subsection{Example: calculating \texorpdfstring{$F_{int} - F_s$}{Fint - Fs} for screened Coulomb potential}
\label{subsec:coulomb_potential}

In section~\ref{sec:dust_crystal} grains in dusty plasma will be considered and in section~\ref{sec:electrons_on_helium} we'll set our eyes on electrons on the liquid-helium surface.
In both cases we have to deal with Coulomb potential\,--- screened for dust particles and regular for electrons.
The latter will be treated as screened Coulomb with screening distance approaching infinity.
During this example we aim to find $F_{int} - F_s$ for screened Coulomb potential using methods of section~\ref{sec:lattice}.
Consideration of physics of this two systems will be postponed until sections~\ref{sec:dust_crystal} and~\ref{sec:electrons_on_helium}.

Screened Coulomb potential can be written as
\begin{equation}
\label{screened_coulomb}
 V(r) = \frac{q^2 e^{-r/\lambda_D}}{r},
\end{equation}
where $\lambda_D$ is Debye screening distance.
It may be shown (appendix \ref{subsec:appendix_coulomb_v_k}) that expression for $V_{\vec{k}}$ should be
\begin{equation}
\label{v_k_coulomb}
 V_{\vec{k}} = \frac{4\pi\bar{\rho}q^2}{1/\lambda_D^2 + 4\pi^2\vec{k}^{\,T}\hat{G}\hat{G}^{T}\vec{k}}.
\end{equation}

At this point we already have $V_{\vec{k}}$~\eqref{v_k_coulomb} and $\rho_{\vec{k}}$~\eqref{rho_k}.
So we can write $F_{int}$ from equation~\eqref{f_int_prime}
\begin{equation}
\label{interaction_energy}
 F_{int} = \sum_{\vec{k}\in\mathbb{Z}^3} e^{-4\pi^2s^2 \vec{k}^{\,T} \hat{G} \hat{G}^{T} \vec{k}} 
           \frac{4\pi\bar{\rho}q^2}{1/\lambda_D^2 + 4 \pi^2 \vec{k}^{\,T} \hat{G} \hat{G}^{T} \vec{k}}.
\end{equation}
Compared to straight-forward method of computing, this one has better convergence.
For big values of $\lambda_D$ we will see, that first summands in~\eqref{interaction_energy} descend as $1/|\vec{k}|^2$.
But if we write series containing interaction energy between this particle and any other, first summands will descend only as $1/|\vec{k}|$.

In section~\ref{sec:lattice} we have mentioned that there is one compensatory term we need to take into account.
This term will exclude Coulomb self-interaction.
If we omit technical details (see appendix~\ref{subsec:appendix_f_s}) explicit form of $F_s$ can be obtained
\begin{equation}
\label{f_s_general_potential}
 F_{s} = \frac{1}{2\sqrt{\pi} s^3}
         \int\limits_0^\infty
             e^{-r^{\prime\,2}/(4s^2)}
             V(r') r^{\prime\,2}
         dr',
\end{equation}
or substituting screened Coulomb potential~\eqref{screened_coulomb} as $V$ and approximating for the case when $s$ is much smaller then screening distance $\lambda_D$
\begin{equation}
\label{f_s_coulomb}
    F_s = \frac{q^2}{\sqrt{\pi}s} - \frac{q^2}{\lambda_D}.
\end{equation}

Expressions we've got for $F_{int}$ and $F_s$ are a bit inconvenient.
So we use that $s$ is small compared to distances in the lattice (``sharp peak'' approximation, section~\ref{sec:lattice}) and perform some approximation.

First of all we introduce mean distance calculated from particle's density
\begin{equation}
\label{mean_lattice_dist}
   l = \frac{1}{\sqrt[3]{\bar{\rho}}},
\end{equation}
and split sum \eqref{interaction_energy} into two parts.
Further we will suppose lattice under consideration as not too much degenerated.
In other words, we will assume that it may be described as some ``deformation'' of cubic lattice (appendix \ref{subsec:approximation_f_int})
\begin{equation}
\label{approximate_free_energy}
\begin{aligned}
 F_{int} &- F_{s} = \frac{\sqrt{\pi} q^2}{l}
                    \sum_{\vec{k} \neq 0}
                        \frac{e^{- \vec{k}^{\,T} \hat{G}^{\,-1\,T} \hat{G}^{-1} \vec{k} / l^2}}
                             {\vec{k}^{\,T} \hat{G}^{\,-1\,T} \hat{G}^{-1} \vec{k} / l^2}
+\\
                 &+ \frac{q^2}{\pi l}
                    \sum_{\vec{k} \neq 0}  
                        \frac{e^{ -l^2 \vec{k}^{\,T} \hat{G} \hat{G}^{T} \vec{k}} }
                             {l^2 \vec{k}^{\,T} \hat{G} \hat{G}^{T} \vec{k}}
                  - \frac{2 \sqrt{\pi} q^2}{l} 
                  + 4\pi\bar{\rho}q^2\lambda_D^2.
\end{aligned}
\end{equation}

One may see that last term is equal to energy in case of evenly distributed charge.
Other terms are corrections due to charge pointness and comprise all information on lattice geometry.

It is very useful to notice, that this is the only summand that approaches infinity if $\lambda_D \to \infty$.
This means that \eqref{approximate_free_energy} can be used for different lattice comparison even without screening.
We will only need to measure energy starting from $4\pi\bar{\rho}q^2\lambda_D^2$ as ``ground-level''.

One more thing to notice.
Coulomb is a long-range interaction and localization distance $s$ should not play any significant role in approximation.
Here we see that it is really absent.
Rather different may be a picture for short-range interactions or effective interactions with uncompensated $F_s$ (example~\ref{subsec:capillary_interaction}).

\subsection{Example: calculating \texorpdfstring{$F_{int}$}{Fint} for capillary interaction}
\label{subsec:capillary_interaction}

Last example we will consider is a rough estimation of capillary interaction energy.
This results will be used in section~\ref{sec:electrons_on_helium}.
More discussion on physics of the phenomenon and references we postpone until~\ref{sec:electrons_on_helium}.

It is known, that electrons deform liquid-helium surface when pressed against with electric field and this causes effective interaction with potential~\cite{haque}
\begin{equation}
    V(r) = - \frac{q^2 E^2}{2\pi\sigma}  \textrm{K}_0(r/l_0),
\label{deformPotential}
\end{equation}
where $qE$ is force caused by electric field, $\sigma$\,--- surface tension of liquid helium, $r$\,--- distance between particles, $\text{K}_0$\,--- modified Bessel function, and $l_0 = \sqrt{\sigma/g\rho_{He}}$~\cite{haque}\,--- capillary length that depends on the fluid properties only.

Since we know that capillary interaction is essentially two-dimensional (particles should be localized on some surface, no three-dimensional analogue) we will use~\eqref{f_int} with~\eqref{two_dimensional_matrix} and~\eqref{two_dimensional_rho_k}.
Performing calculations (appendix~\ref{subsec:appendix_capillary_v_k}) we get
\begin{equation}
\label{capillary_v_k}
    V_{\vec{k}} = - \frac{q^2E^2}{\sigma} \frac{\bar{\rho}}{1/l_0^2 + 4\pi^2 \vec{k}^{\,T} \hat{\mathcal{G}} \hat{\mathcal{G}}^{T} \vec{k}}.
\end{equation}
So we can immediately write
\begin{equation}
\label{f_int_cap}
F_{int} = -\sum_{\vec{k}\in\mathbb{Z}^2}  
           \frac{\bar{\rho}q^2E^2 e^{-4\pi^2s^2 \vec{k}^{\,T} \hat{\mathcal{G}} \hat{\mathcal{G}}^{T} \vec{k}} }
                {\sigma \left(1/l_0^2 + 4 \pi^2 \vec{k}^{\,T} \hat{\mathcal{G}} \hat{\mathcal{G}}^{T} \vec{k}\right)}.
\end{equation}

For our purposes (see section \ref{sec:electrons_on_helium}) a very rough approximation would be enough.
So we use the same trick as in appendix \ref{subsec:approximation_f_int}, but perform integration instead of summation and dealing with theta functions (see appendix \ref{subsec:approximation_f_cap})
\begin{equation}
\label{f_int_cap_final}
 F_{int} =  \frac{q^2 E^2}{4\pi\sigma}
            \ln\left(s/l_0\right).
\end{equation}

One should notice, that $s \ll l_0$, so logarithm value is less then zero and thus whole expression \eqref{f_int_cap_final} is less then zero.
It coincides with the fact that capillary interaction makes particles attract each other \cite{skachko,lambert,haque}.

\section{Grains in dusty plasma, dust crystal and its lattice}
\label{sec:dust_crystal}

It is known that grains in dusty plasma interact and even show self-organization in form of melting and crystallization of dust crystalls \cite{klumov_joyce,klumov_morfill,lev_zagorodny,lev_tymchyshyn_cond_matt,thomas_morfill}.

We aim to apply methods developed in section \ref{sec:lattice} to this system.
As a first approximation we suppose that grains interact only as charged particles (screened Coulomb potential).
Obtained equation \eqref{approximate_free_energy} (example \ref{subsec:coulomb_potential}) allows to calculate energy of any lattice just inserting correct matrix $\hat{G}$ and performing summation.
But much more interesting is finding the lattice with minimal energy when particles density $\bar{\rho}$ is fixed.
Obtained result can be verified experimentally.

Both sums in \eqref{approximate_free_energy} are highly convergent and this is very convenient for numeric calculations.
Thus we will limit summation with $|\vec{k}| \leq 4$.
This limit was found experimentally: taking more summands does not change calculated minimal lattice and free energy.
On the other hand, taking less results in wrong lattice.

And one more thing to mension.
Searching for lattice with minimal energy does not need calculation of full expression for $F_{int} - F_s$.
We need to minimize only parts with $\hat{G}$.
Moreover, they both contain $q/l$, which means, that results of minimization do not depend of particle charge.

Minimization was performed with standard function FindMinimum of Wolfram Mathematica v.9.
We use \eqref{approximate_free_energy} as function to be minimized.
Particles density is fixed in a following way.
We consider
\begin{equation}
\label{eq_mean_dist}
    \frac{1}{l^3} = \bar{\rho} = \frac{1}{abc\sin(\alpha)\sin(\beta_c)},
\end{equation}
suppose $a = l(1 + \delta a)$, $b = l(1 + \delta b)$ and express $c$ through other parameters to keep constant charge density.
Performing this one may notice that $l$ cancel out in exponents and denominators \eqref{approximate_free_energy}.

Before considering results, one more remark should be done.
Mapping between set of all possible parameter values (translation vectors) and all possible Bravais lattices is not a bijection.
It is rather a surjection\,--- one lattice may be described with different sets of parameters (different translation vectors).
For minimization algorithms to be not confused, we provide some restrictions.

\begin{figure}
\centering
\includegraphics[width=6cm]{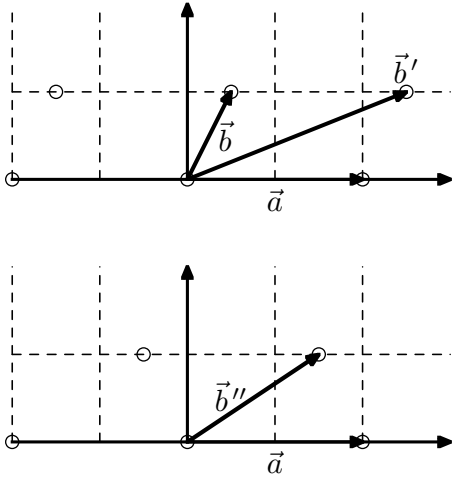}
\caption{
Same lattice in different representations.
It has translation symmetry with respect to two vectors $\vec{a}$ and $\vec{b}$.
Upper figure shows different choice for second translation vector $\vec{b}$ and $\vec{b}^{\,\prime}$.
Using reflections one may show that lattice on bottom figure is the same as on the upper one.
So $\vec{b}^{\,\prime\prime}$ is possible choice as well.
When performing numerical calculations additional restrictions are used to overcome this uncertainty.
}\label{fig:lattice_uncertainty}
\end{figure}

Figure \ref{fig:lattice_uncertainty} shows uncertainty when choosing parameters for lattice description for 2D case.
One may see that it can be eliminated if we claim that projection of $\vec{b}$ on $\vec{a}$ is less then a half of $\vec{a}$.
In terms of parameters this may be expressed as
\begin{equation}
\label{restriction_b}
    0 \leq (1 + \delta b) \cos(\alpha) \leq \frac{1 + \delta a}{2}.
\end{equation}

Restrictions for 3D case are more sophisticated.
But they may be achieved similarly as \eqref{restriction_b}.
Actually \eqref{restriction_b} should be kept for 3D as is.
We only need to add analogous considerations for $\vec{c}$.

\begin{figure}
\centering
\includegraphics[width=8cm]{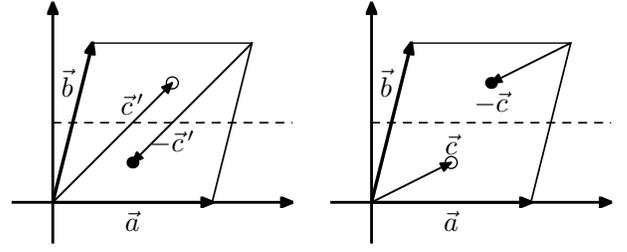}
\caption{
Projection of the same lattice in different representations on $XY$ plane.
Particle from layer above $XY$ plane is shown with circle and below the plane with filled circle.
Particles in $XY$ plane are not explicitly shown.
If particle from above the plane is projected into the upper half of $ab$ parallelogram a reflection with respect to $XY$ plane can be used to get representation where this particle is in lower half of parallelogram.
}\label{fig:lattice_uncertainty_3D}
\end{figure}

First of all we introduce projection of vector $\vec{c}$ on $XY$ plane (see figure~\ref{fig:vectors}) and take particles density into account (substituting $c$ from~\eqref{eq_mean_dist})
$$
\begin{aligned}
    c_x &= c \cos(\beta_c) \cos(\alpha_c) = \frac{l \cot(\beta_c) \cos(\alpha_c)}{(1 + \delta a)(1 + \delta b)\sin(\alpha)},\\
    c_y &= c \cos(\beta_c) \sin(\alpha_c) = \frac{l \cot(\beta_c) \sin(\alpha_c)}{(1 + \delta a)(1 + \delta b)\sin(\alpha)}.
\end{aligned}
$$
Obviously this projection should fall into parallelogram constructed on $\vec{a}$ and $\vec{b}$.
This condition is an obvious extension of two-dimensional case to third dimension.
But we can strenghten this condition by using reflection with respect to $XY$ plane (figure~\ref{fig:lattice_uncertainty_3D}).
We can always assume that $\vec{c}$ is projected into lower part of parallelogram constructed with $\vec{a}$ and $\vec{b}$.

As result we get following conditions
$$
\begin{aligned}
    0                          &< \frac{c_y}{l} < \frac{(1 + \delta b) \sin(\alpha)}{2}\\
    \frac{c_y \cot(\alpha)}{l} &< \frac{c_x}{l} < \frac{c_y \cot(\alpha)}{l} + (1 + \delta a)\\
\end{aligned}
$$

\begin{figure}[htb]
  \centering
  \includegraphics[width = 5cm]{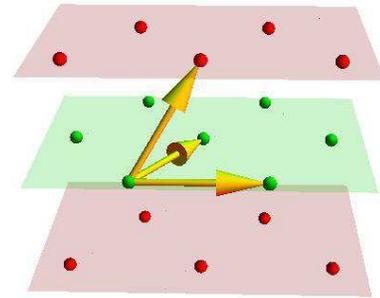}
  \caption{
Part of lattice obtained from minimization.
As is seen, 3-D lattice consists of interleaving planes with 2-D hexagonal (triangular) lattices in each.
The only difference between planes is shift: particles from one plane project onto ``empty spaces'' in another.
Translation vectors are shown as well.
  }\label{figure_hcp}
\end{figure}

Now we can present results of numerical minimization in form of figure~\ref{figure_hcp} and following table.

Looking at figure~\ref{figure_hcp}, one my see, it is very similar to hexagonal close packing (HCP) lattice.
HCP consists of two interleaving planes with triangular 2-D lattice in each as well.
So we may check, if this is really HCP lattice by comparing their parameters.
\begin{center}
\begin{tabular}{|c|c|c|}
\hline
Parameter  & Calculated         & Hexagonal close                                     \\
           & value              & packing (HCP)                                       \\
\hline
$\alpha$   & $\approx 1.04717$  & $\pi/3 \approx 1.0472$                              \\
$\alpha_c$ & $\approx 0.523589$ & $\pi/6 \approx 0.5236$                              \\
$\beta_c$  & $\approx 0.955205$ & $\arcsin\left(\!\!\sqrt{2/3}\right) \approx 0.9553$ \\
$a$        & $\approx 1.122462$ & $\sqrt[6]{2} \approx 1.12246$                       \\
$b$        & $\approx 1.122462$ & $\sqrt[6]{2} \approx 1.12246$                       \\
\hline
\end{tabular}
\end{center}

Results show that lattice obtained by numerical calculations is really HCP.
This complies with experimental data and computer simulation \cite{klumov_joyce,klumov_morfill}.

\section{Electrons on the liquid-helium surface and their localization distance}
\label{sec:electrons_on_helium}

In this section we consider electrons on the liquid helium surface, calculate localization distance and compare it to obtained in other works.

In the presence of an external field electrons can be pressed against the helium surface with significant force.
But they cannot go through this surface, because quantum effects push them out \cite{lambert,skachko,edelman}.
Thus they deform surface and this deformation, in its turn, changes interaction potential between electrons.
The problem of finding explicit expression of this potential is solved by adding to Coulomb interaction effective capillary interaction.
The lateral for two electrons on the helium surface was calculated in~\cite{haque}.

We aim to find localization distance for electrons on the liquid helium surface.
To achieve this goal we will minimize single-particle's free energy $F_{\text{sp}}$ over $s$ (localization distance, see example \ref{subsec:two_dim_fermi}).
So first of all we construct $F_{\text{sp}}$ and find $\partial F_{\text{sp}} / \partial s$.

In section \ref{sec:stat_descr} we developed a method for treating some quantum systems.
Example \ref{subsec:two_dim_fermi} shows how to use it for two-dimensional system of Fermi-particles.
So we consider $F_{ent}$ from \eqref{entropy_part} as free energy of one particle caused with entropy in the system.

Potential energy is more complicated.
In section \ref{sec:lattice} we developed methods for its calculation.
There are two parts of potential energy for electrons on the liquid helium surface \cite{lambert,skachko,edelman,haque}: Coulomb interaction and capillary interaction.
For Coulomb interaction we can find a shortcut\,--- it was shown in \cite{lev_tymchyshyn} that this part approximately does not depend on localization distance $s$.
Besides, one may think on approximate expression for Coulomb interaction from example \ref{subsec:coulomb_potential} \eqref{approximate_free_energy}.
It does not contain $s$, and the same behavior we may expect from Coulomb interaction for 2D lattice.
Since we are interested in derivative $\partial F_{\text{sp}} / \partial s$, consideration of this part of potential energy may be omitted.
More interesting is capillary interaction presented in example \ref{subsec:capillary_interaction}, approximation \eqref{f_int_cap_final}.
It contains explicit dependence on $s$ and thus should be taken into account.

Bringing it all together we get
$$
   \frac{\partial F_{\text{sp}}}{\partial s} = 
 - \frac{\lambda_T^2}{4 \pi^2 \beta s^3}
 - \frac{\pi^4 s}{3\beta\lambda_T^2}
 + \frac{q^2 E^2}{4\sigma\pi s},
$$
independently on lattice we are considering.

Substituting electron charge $q \rightarrow \bar{e}$ and solving equation $\partial F / \partial s = 0$ we get
\begin{equation}
\label{general_s}
 s^2 = \frac{3 \hbar^2 \beta^2 \bar{e}^2 E^2}{4 \pi^3 \sigma m} 
       \left( 1 - \sqrt{ 1 - \frac{16 \sigma^2 \pi^4}{3 \bar{e}^4 E^4 \beta^2} } \right)
\end{equation}
If temperature tends to be small $T \to 0$ (e.g. $\beta \to \infty$)
\begin{equation}
\label{classic_s}
 s = \sqrt{\frac{2\pi\sigma\hbar^2}{m\bar{e}^2E^2}}.
\end{equation}
This result does not differ from the classical one presented in~\cite{edelman}.

If we consider \eqref{general_s} more carefully, some interesting result may be obtained.
Let's consider low densities so that any value of $s$ may be treated as much lower than interelectron distance $l$.
Then we may notice that solution for $s$ exists if
\begin{equation}
\label{lattice_existence}
 \frac{E^2}{T} \geq \frac{4 \sigma \pi^2 k}{\sqrt{3} \bar{e}^2} \approx 2100\,\frac{dyn}{cm \cdot K}.
\end{equation}

Last equation means that there is no lattice if temperature $T$ is too high, or if field $E$ is too weak.
This statement is in agreement with our physical intuition.

\section{Conclusions}
\label{sec:conclusions}

In conclusion we want to make some overview of results we have obtained.

In \textbf{section~\ref{sec:stat_descr}} we started with a statistical description of inhomogeneous system of interacting particles.
It was shown that variety of quantum systems may be described with no much difference from classical ones.
We found an expression for free energy \eqref{free_energy}.
The main feature of this equation appeares in expression for ``entropy part'' of free energy.
Obtained result is applicable to a wide variety of systems with different number of dimensions.

To make a step aside from pure theory we considered an example~\ref{subsec:two_dim_fermi} of two-dimensional system of Fermi particles.
This result was later reused in section~\ref{sec:electrons_on_helium}.

In \textbf{section~\ref{sec:lattice}} we stated that particles are arranged in a lattice.
It seems, that adding translation symmetry to consideration we can get some simplification for potential energy expressions \eqref{PDF_in_Fourier} and \eqref{f_int}.
One may notice that for any ``catastrophic'' potential there is only one diverging term.
But this term does not contain information on ``lattice geometry'', it depends on mean particle density only.
This means, we can compare two lattices with equal mean particle density even if inter-particle potential is ``catastrophic''.

In this approach all potentials are classified into two groups: ones with self-interaction (often effective interactions through medium) and without.
Examples \ref{subsec:coulomb_potential} and \ref{subsec:capillary_interaction} demonstrate how to use proposed technique in both cases.
Later they are reused for treating grains in dusty plasma (section~\ref{sec:dust_crystal}) and electrons on the liquid-helium surface (section~\ref{sec:electrons_on_helium}).

In \textbf{section~\ref{sec:dust_crystal}} we considered grains in dusty plasma.
We started with potential energy obtained in examle \ref{subsec:coulomb_potential} and minimized it over all possible lattices.
Since there are no good approximations for theta-function it was too complicated to perform this minimization analytically.
Thus we used numerical computations.
This computations have some features presented in a real-life case \ref{sec:dust_crystal}.
Results obtained in \ref{sec:dust_crystal} show that we should expect hexagonal close packing (HCP) lattice for grains in dusty plasma.
This coincides with numerical simulations in \cite{klumov_morfill} and experiments \cite{klumov_joyce}.

Last \textbf{section~\ref{sec:electrons_on_helium}} shows how to use developed methods for electrons on the liquid-helium surface.
It is shown that in some cases much more simple (but more rough as well) compared to section \ref{sec:dust_crystal} approximation may be obtained.
This approximation is used to calculate localization distance and then compared to classical result \cite{edelman}.
When temperature $T \to 0$ they fully coincide.
Besides, basing on this results some ``guess'' when lattice can exist is done.
Obtained equation can be verified experimentally.

\section{Appendices}
\label{sec:appendices}

During this article we tried to provide rigorous exposition of presented ideas, but without physical sole being lost in lots of equations.
Thus, calculations that heavily rely on mathematical transfomations only, were moved to appendices.
Reader, interested only in ``physical part'', may omit them.

\subsection{Integration over momentum in \texorpdfstring{$\beta F[\mu]$}{F[m]}.}
\label{subsec:appendix_momentum_integration}

We start with \eqref{f_mu_nonintegrated}.
First two summands can be easily integrated if we mention
$$
\frac{1}{e^x+1} = \frac{e^{-x}}{1+e^{-x}},
$$
but last summand should be integrated in terms of special functions, namely dilogarithm~\cite{bateman_1}
$$
\text{Li}_2(z) = - \int\limits_0^z \frac{\ln(1-z)}{z} dz.
$$
If we mention that
$$
\frac{1}{1+e^x} = 1 - \frac{1}{1+e^{-x}}.
$$
we would get
\begin{equation}
  \begin{aligned}
\beta F[\mu] &=      \frac{m^2}{8 \pi^2 \hbar^4 \beta} \iint d^2 \vec{r}\, d^2 \vec{r}^{\,\prime} V(|\vec{r}-\vec{r}^{\,\prime}|) \times\\
             &\times \ln\left(1+e^{\beta \mu(\vec{r}\,)}\right) \ln\left(1+e^{\beta \mu(\vec{r}^{\,\prime})}\right) +\\ 
             &+      \frac{m}{2 \pi \hbar^2 \beta} \int d^2 r
                     \left[ 
                         \beta\mu(\vec{r}\,)\ln\left(1+e^{\beta \mu(\vec{r}\,)}\right) -\right.\\
             &\left.   - \frac{\beta^2\mu^2(\vec{r}\,)}{2}
                       - \text{Li}_2 \left(-e^{-\beta \mu (\vec{r}\,)} \right) - \frac{\pi^2}{6} 
                     \right],
  \end{aligned}
\end{equation}
which can be simplified further.

First of all we will use Landen's identity~\cite{bateman_1}
\begin{equation}
\begin{aligned}
 &\text{Li}_2\left(1-z\right) + \text{Li}_2\left(1-\frac{1}{z}\right) = - \frac{1}{2} \ln^2(z),\\
 &z\in\mathbb{C} \setminus \left]-\infty;0\right] 
\end{aligned}
\end{equation}
with $z = 1 + e^{-\beta\mu(\vec{r}\,)}$ and expand $\ln^2\left(1+e^{-\beta\mu(\vec{r}\,)}\right)$.
This leads to
$$
\begin{aligned}
\beta F[\mu] &=      \frac{m^2}{8 \pi^2 \hbar^4 \beta} \int \int d^2 \vec{r}\, d^2 \vec{r}^{\,\prime} V(|\vec{r}-\vec{r}^{\,\prime}|) \times\\
             &\times \ln\left(1+e^{\beta \mu(\vec{r}\,)}\right) \ln\left(1+e^{\beta \mu(\vec{r}^{\,\prime})}\right) +\\ 
             &+      \frac{m}{2 \pi \hbar^2 \beta} \int d^2 r
                     \left[ 
                         \text{Li}_2\left(\frac{1}{1+e^{\beta\mu(\vec{r}\,)}}\right) - \frac{\pi^2}{6} +\right.\\
             &\left.   + \frac{1}{2}\ln^2\left(1+e^{\beta\mu(\vec{r}\,)}\right)
                     \right].
\end{aligned}
$$

Now we substitute~\eqref{rho_to_mu} and~\eqref{lambda_t} into the last expression and get~\eqref{free_energy_final}.

\subsection{On the properties of \texorpdfstring{$\mathfrak{f}_{\vec{k}}(\vec{r}\,)$}{f(r)}.}
\label{subsec:appendix_properties_of_f_k}

In \eqref{PDF_in_Fourier} we present how to calculate probability distribution function for the whole system basing on single-particle probability distribution function.
It is based on the decomposition of $\rho_{\text{sp}}$ using $\mathfrak{f}_{\vec{k}}(\vec{r}\,)$ \eqref{PDF_in_Fourier_f_k} as basis functions in inverse lattice space.
Here we add some consideration about connection between Fourier series and presented one.

Suppose we have a function $g(\vec{r}\,)$ defined on the unit cube $\mathbb{U} \equiv [0;1] \times [0;1] \times [0;1]$.
This function can be expressed in terms of Fourier series
\begin{equation}
\label{appendix_fourier}
\begin{aligned}
 g(\vec{r}\,) &= \sum\limits_{\vec{k}\in\mathbb{Z}^3} g_{\vec{k}}\, e^{2\pi\mathbf{i}\left(\vec{k}^{\,T}\vec{r}\,\right)}, \\
 g_{\vec{k}}  &= \iiint\limits_{\mathbb{U}} g(\vec{r}\,) e^{-2\pi\mathbf{i}\left(\vec{k}^{\,T}\vec{r}\,\right)} d^3\vec{r}.
\end{aligned}
\end{equation}

Let's consider a bijection $\hat{G}$ \eqref{PDF_in_Fourier_G_operator} from $\mathbb{V}$ to $\mathbb{U}$ and its inverse $\hat{G}^{-1}$ (see Fig.~\ref{fig:vectors} for geometrical reasoning)
\begin{equation}
\label{appendix_g_inverse}
 \hat{G}^{-1} = \begin{pmatrix}
                 a & b \cos(\alpha) & c \cos(\alpha_c)\cos(\beta_c) \\
                 0 & b \sin(\alpha) & c \sin(\alpha_c)\cos(\beta_c) \\
                 0 & 0              & c \sin(\beta_c)
                \end{pmatrix}
\end{equation}
It may be seen that rows of $\hat{G}$ are components of inverse lattice basis vectors.
We are interested in case $g(\vec{r}\,) = \rho_{\text{sp}}\left(\hat{G}^{-1}\vec{r}\,\right)$.
Since we know that $\forall \vec{r} \in \mathbb{V}: \rho_{\text{sp}}\left(\hat{G}^{-1}\hat{G}\vec{r}\,\right) = \rho_{\text{sp}}(\vec{r}\,)$ one can immediately write
\begin{equation}
\label{appendix_f1}
 \rho_{\text{sp}}(\vec{r}\,) = \sum\limits_{\vec{k}\in\mathbb{Z}^3} g_{\vec{k}}\, e^{2\pi\mathbf{i}\left(\vec{k}^{\,T}\hat{G}\vec{r}\,\right)}.
\end{equation}
And same way we consider second equation from \eqref{appendix_fourier}
\begin{equation}
\label{appendix_f2}
 g_{\vec{k}} = \iiint\limits_{\mathbb{V}} \rho_{\text{sp}}\left(\hat{G}^{-1}\vec{r}\right) 
               e^{-2\pi\mathbf{i}\vec{k}^{\,T}\hat{G}\hat{G}^{-1}\vec{r}} 
               \frac{d^3\left(\hat{G}^{-1}\vec{r}\right)}{J \left[\hat{G}^{-1}\vec{r}\,\right]},
\end{equation}
where $J$ is Jacobian.
Here $\hat{G}^{-1}\vec{r}$ is treated as new variable with domain $\mathbb{V}$.

Jacobian $J \left[\hat{G}^{-1}\vec{r}\,\right] = abc\sin(\alpha)\sin(\beta_c) = 1/\bar{\rho}$ is actually the volume of $\mathbb{V}$ or inverse mean particle density.
Designating $g_{\vec{k}}$ as $\rho_{\vec{k}}$ in~\eqref{appendix_f1} and \eqref{appendix_f2}, changing variable in~\eqref{appendix_f2} $\hat{G}^{-1}\vec{r} \rightarrow \vec{r}$ and designating exponent in~\eqref{appendix_f1} as $\mathfrak{f}_{\vec{k}}$ (it can be seen that exponent in~\eqref{appendix_f2} is $\mathfrak{f}_{\vec{k}}^{*}$, where $*$ designates complex conjugation) we immediately get equations \eqref{PDF_in_Fourier}.
Moreover, obtained result means that all properties of Fourier series can be applied to decomposition \eqref{PDF_in_Fourier}.

Last thing to mention are periodical properties of presented series.
We are going to expand domain to $\mathbb{R}^3$ so that $\rho(\vec{r}\,)$ will be defined everywhere in the space.
So now we need to explore behavior of this function.
From \eqref{PDF_in_Fourier_G_operator} one may see, if $l \in \mathbb{Z}$, $m \in \mathbb{Z}$ and $n \in \mathbb{Z}$:
\begin{equation}
 \hat{G}\left(\vec{r} + l\vec{a} + m\vec{b} + n\vec{c}\,\right) = \hat{G}\vec{r} + l\vec{e}_x + m\vec{e}_y + n\vec{e}_z,
\end{equation}
where $\vec{e}_x$, $\vec{e}_y$ and $\vec{e}_z$ are unit vectors along coordinate axis.
With regard to \eqref{PDF_in_Fourier_f_k}
\begin{equation}
 \mathfrak{f}_{\vec{k}}\left(\vec{r} + l\vec{a} + m\vec{b} + n\vec{c}\,\right) = \mathfrak{f}_{\vec{k}}(\vec{r}\,)
\end{equation}
and from \eqref{PDF_in_Fourier_rho_xy}
\begin{equation}
 \rho\left(\vec{r} + l\vec{a} + m\vec{b} + n\vec{c}\,\right) = \rho(\vec{r}\,).
\end{equation}
Last equation justifies our view of connection between $\rho$ and $\rho_{\text{sp}}$ as it is presented on Fig.\,\ref{fig:one_to_many}.

\subsection{On the expression of \texorpdfstring{$\rho_{\vec{k}}$}{rho k} for Gaussian \texorpdfstring{$\rho_{\text{sp}}$}{rho sp}}
\label{subsec:appendix_rho_k}

We suppose that $\rho_{\text{sp}}(\vec{r}\,)$ is either $0$ everywhere in $\mathbb{R}^3 \setminus \mathbb{V}$, or at least negligibly small.
If so, we can change integration limits in~\eqref{PDF_in_Fourier_rho_k} to infinite.

At this point we are interested in specific form of $\rho_{\text{sp}}$ \eqref{rho_sp}, thus it is explicitly substituted into~\eqref{PDF_in_Fourier_rho_k}.
Besides we designate $\vec{\mathcal{K}} = 2\pi \vec{k}^{\,T} \hat{G}$ and rewrite expression in Cartesian coordinates changing multiple integral to the product of integrals
$$
    \rho_{\vec{k}} 
  = \frac{\bar{\rho}}{(2 \pi s^2)^{3/2}}
    \prod_{j = 1}^{3}
    \int\limits_{-\infty}^{\infty}
            e^{-r_j^2/(2s^2) -\mathbf{i}\mathcal{K}_j r_j }
    d r_j.
$$

Last expression can be integrated if we use following relation~\cite{gradstein_ryzhyk}
$$
    \int\limits_{-\infty}^{+\infty} e^{-p^2x^2 \pm qx} dx = \frac{\sqrt{\pi}}{p} e^{q^2/(2p)^2},\quad \Re\left(p^2\right) > 0.
$$
As result we will get
$$
    \rho_{\vec{k}} 
  = \bar{\rho}
    \prod_{j = 1}^{3}
        e^{-\mathcal{K}_j^2 s^2 / 2}
    d r_j.
$$

Changing product of exponents to exponent of sum and mentioning that $\sum_j \mathcal{K}_j^2 = 4\pi^2 \vec{k}^{\,T} \hat{G} \hat{G}^{\,T} \vec{k}$ we immediately get~\eqref{rho_k}.

\subsection{On the expression of \texorpdfstring{$F_{int}$}{F int}.}
\label{subsec:appendix_expression_f_int}

Lets start with~\eqref{freeEnergySeparated-int}.
Inner integral has infinite borders so we may rewrite this expression as follows
\begin{equation}
\label{appendix_f_int_varchange}
  F_{int} = \iiint\limits_{\mathbb{V}} \iiint\limits_{\mathbb{R}^3}
                V\left(\left|\vec{r}^{\,\prime}\right|\right)
                \rho\left(\vec{r}\,\right) \rho\left(\vec{r}+\vec{r}^{\,\prime}\right) 
            d^3 \vec{r}^{\,\prime}\, d^3 \vec{r}.
\end{equation}
Since $\rho\left(\vec{r}\,\right)$ is real it can be replaced with complex conjugate $\rho^{*}\left(\vec{r}\,\right)$ without changing $F_{int}$.
Now we can substitute $\rho$ from \eqref{PDF_in_Fourier_rho_xy} and mention that $\mathfrak{f}_{\vec{k}}\left(\vec{r}+\vec{r}^{\,\prime}\right) = \mathfrak{f}_{\vec{k}}\left(\vec{r}\,\right) \mathfrak{f}_{\vec{k}}\left(\vec{r}^{\,\prime}\right)$ \eqref{PDF_in_Fourier_f_k}.
So if we designate $V_{\vec{k}}$ as in \eqref{f_int_v_k}
it can be written
$$
  F_{int} = \frac{1}{\bar{\rho}}
            \sum\limits_{\vec{k}\in\mathbb{Z}^3} \rho_{\vec{k}} V_{\vec{k}}
            \sum\limits_{\vec{k}^{\,\prime}\in\mathbb{Z}^3} \rho_{\vec{k}^{\,\prime}}^{*}
            \iiint\limits_{\mathbb{V}}
                \mathfrak{f}_{\vec{k}}\left(\vec{r}\,\right) \mathfrak{f}_{\vec{k}^{\,\prime}}^{*}\left(\vec{r}\,\right)
            d^3 \vec{r}.
$$

Last equation can be simplified if we perform integration.
From appendix~\ref{subsec:appendix_properties_of_f_k} we expect orthogonality of $\mathfrak{f}_{\vec{k}}$ functions.
Let's mention from \eqref{PDF_in_Fourier_f_k} $\mathfrak{f}_{\vec{k}}(\vec{r}\,) \mathfrak{f}_{\vec{k}^{\,\prime}}^{*}(\vec{r}\,) = \mathfrak{f}_{\vec{k}-\vec{k}^{\,\prime}}(\vec{r}\,)$ and we will get
\begin{equation}
     \iiint\limits_{\mathbb{V}} \mathfrak{f}_{\vec{k}}(\vec{r}\,) \mathfrak{f}_{\vec{k}^{\,\prime}}^{*}(\vec{r}\,) d^3 \vec{r}
  =  abc\sin(\alpha)\sin(\beta_c) \delta_{\vec{k},\vec{k}^{\,\prime}}.
\end{equation}
Here $\delta_{\vec{k},\vec{k}^{\,\prime}} = \delta_{k_x,k'_x}\delta_{k_y,k'_y}\delta_{k_z,k'_z}$\,--- product of three Kronecker's delta.

Mentioning that $abc\sin(\alpha)\sin(\beta_c) = 1/\bar{\rho}$, where $\bar{\rho}$ is mean particle density (one particle per $\mathbb{V}$, see fig.~\ref{fig:vectors} for geometrical reasoning) we get \eqref{f_int_prime}.

\subsection{\texorpdfstring{$V_{\vec{k}}$}{Vk} for screened Coulomb potential}
\label{subsec:appendix_coulomb_v_k}

Let's consider equations \eqref{PDF_in_Fourier_f_k} and \eqref{PDF_in_Fourier_G_operator} in spherical coordinates
$$
\allowdisplaybreaks
\begin{aligned}
 \mathfrak{f}_{\vec{k}}(\vec{r}\,) &= e^{
                               2\pi\mathbf{i}r
                               \bigl(
                                   g_{\vec{k}}\cos(\theta) + g_{\vec{k}}'\cos(\varphi-\delta\varphi_{\vec{k}})\sin(\theta)
                               \bigr)
                             },\\
 g_{\vec{k}}            &= \frac{k_x\sin(\alpha_c - \alpha)}{a \tan(\beta_c) \sin(\alpha)}
                         - \frac{k_y\sin(\alpha_c)}{b \tan(\beta_c) \sin(\alpha)}
                         + \frac{k_z}{c \sin(\beta_c)},\\
 g_{\vec{k}}'           &= \sqrt{\frac{k_x^2}{a^2} + \left(\frac{k_y}{b \sin(\alpha)} - k_x \frac{\cot(\alpha)}{a}\right)^2},
\end{aligned}
$$
and rewrite $V_{\vec{k}}$ from \eqref{f_int_v_k}
\begin{multline*}
  V_{\vec{k}} = \bar{\rho}
                \int\limits_0^\infty
                    dr V(r) r^2
                    \int\limits_0^{\pi}
                    d\theta\sin(\theta) \times\\
                        \underbrace{\times\int\limits_0^{2\pi}d\varphi
                            e^{2\pi\mathbf{i}r \bigl(g_{\vec{k}}\cos(\theta) + g_{\vec{k}}'\cos(\varphi-\delta\varphi_{\vec{k}})\sin(\theta)\bigr) }
                        }_{I(\theta)}.
\end{multline*}

Integrating over $\varphi$ we get
$$
   I(\theta) = 2 \pi e^{2\pi\mathbf{i} g_{\vec{k}} r \cos(\theta)} \text{J}_0\left(2\pi g_{\vec{k}}' r \sin(\theta)\right).
$$
Then we substitute last expression into equation for $V_{\vec{k}}$
$$
\begin{aligned}
 V_{\vec{k}} &= 2 \pi \bar{\rho}
                \int\limits_0^\infty
                    V(r) r^2
                    \int\limits_0^{\pi}
                        e^{2\pi\mathbf{i} g_{\vec{k}} r \cos(\theta)} \times\\
             &\times    \text{J}_0\left(2\pi g_{\vec{k}}' r \sin(\theta)\right)
                    \sin(\theta)d\theta
                dr.
\end{aligned}
$$

We will consider screened Coulomb potential \eqref{screened_coulomb}, which means integration over $r$ can be performed.
One may use following relation \cite{gradstein_ryzhyk}
\begin{multline*}
    \int\limits_0^\infty
        e^{-\alpha x} \text{J}_\nu(\beta x) x^{\nu + 1}
    dr
 =  \frac{2\alpha (2\beta)^\nu \Gamma\left(\nu + \frac{3}{2}\right)}
         {\sqrt{\pi}(\alpha^2+\beta^2)^{\nu+3/2}                   },\\
    \Re(\alpha) > |\Im(\beta)|, \Re(\nu) > -1
\end{multline*}
and combine it with equation for $V_{\vec{k}}$ to get
$$
  V_{\vec{k}} = 
                \int\limits_0^{\pi}
                    \frac{2 \pi \bar{\rho} q^2 \left(1/\lambda_D - 2\pi\mathbf{i} g_{\vec{k}} \cos(\theta)\right)\sin(\theta)d\theta}
                         {\left(\left(1/\lambda_D - 2\pi\mathbf{i} g_{\vec{k}} \cos(\theta)\right)^2 + \left(2\pi g_{\vec{k}}' \sin(\theta)\right)^2\right)^{3/2}}.
$$
Changing variable $t = \cos(\theta)$ and performing integration over $t$ we will get
$$
   V_{\vec{k}} = \frac{4 \pi \bar{\rho} q^2}{1/\lambda_D^2 + 4\pi^2 g_{\vec{k}}^2 + 4\pi^2 g_{\vec{k}}^{\prime\,2}}.
$$
Last expression is equivalent to \eqref{v_k_coulomb}.
One may check this by expanding $g_{\vec{k}}$ and $g_{\vec{k}}'$.

\subsection{Calculation of \texorpdfstring{$F_s$}{Fs} for screened Coulomb potential.}
\label{subsec:appendix_f_s}

We will start with equation~\eqref{freeEnergySeparated-self} and assume that $\rho_{\text{sp}}$ is a very ``sharp'' function (section~\ref{subsec:two_dim_fermi}).
With this assumption we can change integration limits to infinite and this, in turn, will allow us to perform variable exchange as we did in~\eqref{appendix_f_int_varchange}
\begin{equation}
\label{appendix_general_f_s}
  F_{s} = \iiint\limits_{\mathbb{R}^3} \iiint\limits_{\mathbb{R}^3}
              V(|\vec{r}^{\,\prime}|) \rho_{\text{sp}}(\vec{r}\,) \rho_{\text{sp}}(\vec{r}+\vec{r}^{\,\prime})
          d^3 \vec{r}^{\,\prime}\, d^3 \vec{r}. 
\end{equation}

First of all we perform some mathematical transformations with \eqref{rho_sp} (in Cartesian coordinate system) and write the following
$$
   \rho_{\text{sp}}(\vec{r}\,) \rho_{\text{sp}}(\vec{r}+\vec{r}^{\,\prime})
 = \frac{1}{8 \pi^3 s^6}  
   e^{-\left[\vec{r} + \vec{r}^{\,\prime}/2\right]^2/s^2 - \vec{r}^{\,\prime\,2}/\left(4s^2\right)}.
$$
Now we can substitute this expression into \eqref{appendix_general_f_s} and perform integration over $\vec{r}$.
Writing result in spherical coordinate system and integrating over angles we immediately get \eqref{f_s_general_potential}.

Since we know explicit expression for $V$ \eqref{screened_coulomb} we can substitute it into \eqref{f_s_general_potential} and perform integration using definition of the complementary error function~\cite{bateman_2}
\begin{equation}
\label{erfc_definition}
 \text{erfc}(x) = \frac{2}{\sqrt{\pi}} 
                  \int\limits_x^\infty e^{-t^2} dt.
\end{equation}
As result we will get 
\begin{equation}
\label{exact_f_s}
   F_s = \frac{q^2}{\sqrt{\pi}s}
         \left(1 - \frac{s\sqrt{\pi}}{\lambda_D} e^{s^2/\lambda_D^2} \text{erfc}(s/\lambda_D)\right).
\end{equation}
Since we may obviously suppose that $s$ is very small compared to screening distance, then $s \ll \lambda_D$.
So we can approximate last equation and as result we will get \eqref{f_s_coulomb}.

\subsection{Approximation of \texorpdfstring{$F_{int} - F_{s}$}{Fint-Fs} for screened Coulomb potential}
\label{subsec:approximation_f_int}

We start with expression~\eqref{interaction_energy} for $F_{int}$.
But one may see that it contains expressions $s^2 \hat{G} \hat{G}^{T}$.
Components of $\hat{G} \hat{G}^{T}$ matrix in \eqref{interaction_energy} are proportional to different products of inverse distances in a lattice, e.g. $1/a^2$, $1/(ab)$, and so on (see~\eqref{PDF_in_Fourier_G_operator}).
Since $s \ll a$, $s \ll b$ and $s \ll c$ (assumption about ``sharp Gaussian'') we expect $\left|s^2 \hat{G} \hat{G}^{T}\right| \ll 1$.
This means, that exponent ``starts acting'' only for terms with very large $|\vec{k}|$.
In this section we will try to rewrite equations to get even better convergence then we have.

First of all we rewrite expression for $F_{int}$ as follows
$$
\begin{aligned}
 F_{int} &= \frac{2 \bar{\rho} q^2}{\pi} e^{s^2/\lambda_D^2}\!\!\!
            \int\limits_{2 \pi s}^{\infty}\!\!
                \mathfrak{s}
                e^{-\mathfrak{s}^2/\left(2\pi\lambda_D\right)^2}
                \sum_{\vec{k}\in\mathbb{Z}^3} e^{-\mathfrak{s}^2 \vec{k}^{\,T} \hat{G} \hat{G}^{T} \vec{k}}
            d\mathfrak{s}.
\end{aligned}
$$
One may easily integrate this expression and check that it is equal to \eqref{interaction_energy}.

We notice, that variable $\mathfrak{s}$ takes values much smaller then mean distance~\eqref{mean_lattice_dist} as well as much bigger.
Thus we split this integral into two.
First integration we will provide from $2 \pi s$ to $l$ and second from $l$ to infinity.

We assume that lattice is not too degenerated (section~\ref{subsec:coulomb_potential}), which means that angles $\alpha$ and $\beta_c$ should not significantly differ from $\pi/2$.
So we may expect elements of $\mathfrak{s}^2 \hat{G} \hat{G}^{T}$ to be less than $1$ for first integral and greater than $1$ for second.
If so, then second integration may be performed and we end up with the sum that converges much better than \eqref{interaction_energy}.
$$
\begin{aligned}
 F_{int} &= \frac{2 \bar{\rho} q^2}{\pi} e^{s^2/\lambda_D^2}\!\!
            \int\limits_{2 \pi s}^{l}
                \mathfrak{s}
                e^{-\mathfrak{s}^2/\left(2\pi\lambda_D\right)^2}
                \underbrace{\sum_{\vec{k}\in\mathbb{Z}^3} e^{-\mathfrak{s}^2 \vec{k}^{\,T} \hat{G} \hat{G}^{T} \vec{k}}}_{\Theta\left(0; \mathfrak{s}^2\hat{G}\hat{G}^{T}\right)}
            d\mathfrak{s} +\\
         &+ 4\pi\bar{\rho}q^2 
            \sum_{\vec{k}\in\mathbb{Z}^3}  
                \frac{e^{s^2/\lambda_D^2 - l^2 \left( 1/\left[2\pi\lambda_D\right]^2 + \vec{k}^{\,T} \hat{G} \hat{G}^{T} \vec{k} \right) } }
                     {1/\lambda_D^2 + 4 \pi^2 \vec{k}^{\,T} \hat{G} \hat{G}^{T} \vec{k}}.
\end{aligned}
$$

From \eqref{PDF_in_Fourier_G_operator} one may check with Sylvester's criterion~\cite{bellman_matrix} that $\hat{G}$~\eqref{PDF_in_Fourier_G_operator} is a positive-definite matrix.
Obviously, we expect $\hat{G}^T$ and their product $\hat{G}\hat{G}^{T}$ to be positive-definite matrices as well.
Since $\mathfrak{s}$ takes only positive values $\mathfrak{s}^2 \hat{G} \hat{G}^{T}$ is positive as well.
This means highlighted sum in the last equation is a Riemann theta function~\cite{bellman_theta} at point $z = 0$, so we appropriately designate it with $\Theta\left(0; \mathfrak{s}^2\hat{G}\hat{G}^{T}\right)$.

Using modular transformation~\cite{bellman_theta}
\begin{equation}
\label{modular_transformation}
    \Theta(z;\hat{A}) = \frac{\pi^{d/2}}{\sqrt{\det\hat{A}}}\, \Theta(\hat{A}^{-1} z;\hat{A}^{-1}),
\end{equation}
where $d$ is number of dimensions ($3$ in this case) we get
$$
    \Theta\left(0;\mathfrak{s}^2\hat{G}\hat{G}^{T}\right) = \frac{\pi^{3/2}}{\mathfrak{s}^3 \det\hat{G}}\, \Theta\left(0;\hat{G}^{-1\,T}\hat{G}^{-1}/\mathfrak{s}^2\right).
$$

We used the fact that $\det\hat{G} = \det\hat{G}^{T}$ and that multiplication by $\mathfrak{s}^2$ is equal to multiplication by diagonal matrix which has all elements equal to $\mathfrak{s}^2$.
Explicit expression for $\hat{G}^{-1}$ may be used from \eqref{appendix_g_inverse}.
Besides, we may notice that 
\begin{equation}
\label{det_G}
    \det\hat{G} = \frac{1}{abc\sin(\alpha)\sin(\beta_c)} = \bar{\rho},
\end{equation}
and rewrite expression for $F_{int}$ as follows
$$
\begin{aligned}
 F_{int} &= 2 \sqrt{\pi} q^2 e^{s^2/\lambda_D^2}\!\!\!
            \sum_{\vec{k}\in\mathbb{Z}^3~}
            \underbrace{\int\limits_{2 \pi s}^{l}
                e^{-\mathfrak{s}^2/\left(2\pi\lambda_D\right)^2 - \left| \hat{G}^{-1} \vec{k} \right|^2 / \mathfrak{s}^2}
            \frac{d\mathfrak{s}}{\mathfrak{s}^2}}_{I\left(\vec{k}\right)} +\\
         &+ 4\pi\bar{\rho}q^2 
            \sum_{\vec{k}\in\mathbb{Z}^3}  
                \frac{e^{s^2/\lambda_D^2 - l^2 \left( 1/\left[2\pi\lambda_D\right]^2 + \left| \hat{G}^{T} \vec{k} \right|^2 \right) } }
                     {1/\lambda_D^2 + 4 \pi^2 \vec{k}^{\,T} \hat{G} \hat{G}^{T} \vec{k}}.
\end{aligned}
$$

Now we can perform integration and find $I\left(\vec{k}\right)$.
Since result is very complicated we first introduce two helper functions
$$
\begin{aligned}
    \phi(L)       &= \frac{e^{-L^2 / \left(2\pi\lambda_D\right)^2}}{L}
                   - \frac{1}{2 \sqrt{\pi} \lambda_D}\text{erfc}\left(\frac{L}{2 \pi \lambda_D}\right),\\
    f_{\alpha}(L) &= e^{\alpha\left|\hat{G}^{-1}\vec{k}\right| / (\pi\lambda_D)} 
                     \text{erfc}\left[\frac{\left|\hat{G}^{-1}\vec{k}\right|}{L} 
                   + \frac{\alpha L}{2\pi\lambda_D}\right],
\end{aligned}
$$
and present result in terms of this functions
$$
\begin{aligned}
    I\left(\vec{k} = 0\right)   &= \phi(2 \pi s) - \phi(l),\\
    I\left(\vec{k}\neq 0\right) &= \sqrt{\pi}\left(f_{-1}(l) - f_{-1}(2 \pi s)\right) / \left|4\hat{G}^{-1}\vec{k}\,\right|\\
                                &+ \sqrt{\pi}\left(f_{+1}(l) - f_{+1}(2 \pi s)\right) / \left|4\hat{G}^{-1}\vec{k}\,\right|.
\end{aligned}
$$

Using precise expression for $F_s$~\eqref{exact_f_s} one may see

$$
    F_s = 2 \sqrt{\pi} q^2 e^{s^2/\lambda_D^2} \phi(2 \pi s).
$$
So subtracting $F_s$ from $F_{int}$ simply means neglection of this term.

Our previous calculations do not rely on any approximation, but expression for $I\left(\vec{k}\neq 0\right)$ is very complicated, so we need to perform one.
Before we do, one should show it is acceptable to approximate this sum.
Subsequent calculations are divided into two parts: \textit{uniform convergence proof} and \textit{approximation} itself.

\textbf{Proof of uniform convergence}.
We will use Cauchy criterion for sequence of functions $f_i(x)$ in domain $E$ to achieve this goal: $\forall \varepsilon > 0 ~ \exists N ~ \forall m \geq n > N : \forall x \in E: \left| \sum_{i=n}^m f_i(x) \right| < \varepsilon$.
Further $\sum_{\vec{k}\in\mathbb{Z}^3~} I\left(\vec{k}\right)$ acts as a sequence for convergence checking and entries of $\hat{G}^{-1}$ as well as $s$, $l$ and $\lambda_D$ are treated as variables from domain of admissible parameters.
We have already examined properties of $\hat{G}^{-1}$ and will not get into details again.

First of all we use $s < l < \lambda_D$.
This relation holds for physical system under consideration and we will use it in all subsequent calculations.
This means $f_\alpha(l) > f_\alpha(2 \pi s)$ and thus all terms in sum will be positive and we can avoid using absolute value.
The second thing that appears is following bounding
$$
    0 
      < \sum_{\vec{k}\in\mathbb{Z}^3 \setminus \vec{0}} I\left(\vec{k}\right) 
      < \sum_{\vec{k}\in\mathbb{Z}^3 \setminus \vec{0}} \frac{f_{-1}(l) + f_{+1}(l)}{\left|4\hat{G}^{-1}\vec{k}\,\right|} \sqrt{\pi}.
$$
Then we use relation $\text{erfc}(x) \leq e^{-x^2}$ for $x > 0$ \cite{chiani} and replace $\text{erfc}$ in $f_{\pm 1}(l)$.
Expanding squares in exponents and performing simplification we get
$$
    \sum_{\vec{k}\in\mathbb{Z}^3 \setminus \vec{0}} I\left(\vec{k}\right) 
  < \sum_{\vec{k}\in\mathbb{Z}^3 \setminus \vec{0}} 
    \frac{\exp\left( -\frac{\left|\hat{G}^{-1}\vec{k}\right|^2}{l^2} - \frac{l^2}{4\pi^2\lambda_D^2} \right)}
         {\left|\hat{G}^{-1}\vec{k}\,\right|}.
$$
Besides $2\sqrt{\pi} < 4$ was used.

Now we can use known relation $\min_{\|\vec{x}\|_2=1} \|\hat{A}\vec{x}\|_2 = \sqrt{\lambda_{\text{min}}}$, where $\lambda_{\text{min}}$ is minimal eigenvalue of $\hat{A}^{*}\hat{A}$ and $*$ designates Hermitian adjoint \cite{meyer}.
We know that $\det \hat{G}^{-1} = 1/\bar{\rho} > 0$ and thus $\hat{G}^{-1\,*}\hat{G}^{-1}$ has nonzero eigenvalues.
Taking the minimal one $\lambda_{\text{min}}$ and designating $g = \sqrt{\lambda_{\text{min}}}$ we claim $\left|\hat{G}^{-1}\vec{k}\right| \geq g \left|\vec{k}\right|$.

Now we should change multidimensional summation to one-dimensional.
Since all summands are positive we can rearrange them in any convenient order.
Thus we will perform summation over ``cube surfaces''.
First of all let's designate $\mathbb{K}_n = \{\vec{k} \in \mathbb{Z}^3 |~ |k_x| \leq n \wedge |k_y| \leq n \wedge |k_z| \leq n \}$.
Then total sum can be expressed as
$$
    \sum_{\vec{k}\in\mathbb{Z}^3 \setminus \vec{0}}
    I\left(\vec{k}\right)
  = \sum_{n = 1}^{+\infty}~ \sum_{\vec{k}\in\mathbb{K}_{n} \setminus \mathbb{K}_{n-1}}
    I\left(\vec{k}\right),
$$
where number of summands in $\mathbb{K}_{n} \setminus \mathbb{K}_{n-1}$ can be easily calculated as $(2n+1)^3-(2n-1)^3 \equiv 24n^2+2$.
Minimal length for index vector in this set of indices is $n$.

Collecting together previous calculations we get
$$
    \sum_{\vec{k}\in\mathbb{Z}^3 \setminus \mathbb{K}_{m}} I\left(\vec{k}\right) 
  < \sum_{n = m+1}^{+\infty}
    \frac{(24n^2 + 2) \exp\left( -\frac{g^2 n^2}{l^2} - \frac{l^2}{4\pi^2\lambda_D^2} \right)}{gn}.
$$
Further proof is supposed to be obvious and thus we avoid it.
Proof of uniform convergence for first summand in $F_{int}$ can be performed much more easily by presented scheme and thus is avoided as well.
Instead one useful remark shoud be done.
Series converges better when $g/l$ is bigger.
Taking physical meaning of this variables into account we get following statement: if we project vectors $\vec{a}$, $\vec{b}$ and $\vec{c}$ on axis $X$, $Y$ and $Z$ respectively (see figure~\ref{fig:vectors}) and divide it by mean distance in lattice $\sqrt[3]{\bar{\rho}}$, we will get measure of how good current series converges, or by physical meaning\,--- how close is current lattice to cubic.

\textbf{Approximation.}
We take into account $s \ll l \ll \lambda_D$, neglect all terms containing $l/\lambda_D$ and $s/\lambda_D$ in $\text{erfc}$, and approximate $\text{erfc}(x \to \infty) \sim e^{-x^2}/(x\sqrt{\pi})$ \cite{abramowitz_stegun}.
Obviously $f_\alpha(l) \gg f_\alpha(2 \pi s)$ and thus all $f$ with $2 \pi s$ arguments are neglected.
Besides, we neglect all small terms in exponent and denominators.
To make this procedure more clear we state $\left|\hat{G}^{-1}\vec{k}\right|/l > l/\lambda_D$.
It means screening distance is big enough not to feel lattice deviation from cubic. 
This leads us to equations
$$
\begin{aligned}
    \phi(l)                     &\approx \frac{1}{l},\\
    I\left(\vec{k}\neq 0\right) &\approx \frac{l}{2\left|\hat{G}^{-1}\vec{k}\right|^2}
                                         e^{-\left|\hat{G}^{-1}\vec{k}\right|^2/l^2}.
\end{aligned}
$$
To get the first one, identity $\text{erfc}(x) = 1 - \text{erf}(x)$ and approximation $\text{erf}(x \to 0) \sim 2x/\sqrt{\pi}$~\cite{abramowitz_stegun} were used.

Now approximate equation can be written \eqref{approximate_free_energy}.

\subsection{\texorpdfstring{$V_{\vec{k}}$}{Vk} for capillary interaction}
\label{subsec:appendix_capillary_v_k}

Let's consider two-dimensional version of~\eqref{PDF_in_Fourier_f_k} and \eqref{PDF_in_Fourier_G_operator} in polar coordinates
$$
\allowdisplaybreaks
\begin{aligned}
 \mathfrak{f}_{\vec{k}}(\vec{r}\,) &= e^{
                               2\pi\mathbf{i}r
                               g_{\vec{k}}\cos(\varphi-\delta\varphi_{\vec{k}})
                             },\\
 g_{\vec{k}}                       &= \sqrt{\frac{k_x^2}{a^2} + \left(\frac{k_y}{b \sin(\alpha)} - k_x \frac{\cot(\alpha)}{a}\right)^2},
\end{aligned}
$$
and rewrite $V_{\vec{k}}$ from \eqref{f_int_v_k}
$$
  V_{\vec{k}} = \bar{\rho}
                \int\limits_0^\infty
                dr V(r) r
                \underbrace{\int\limits_0^{2\pi}d\varphi
                    e^{2\pi\mathbf{i}r g_{\vec{k}}\cos(\varphi-\delta\varphi_{\vec{k}}) }
                }_{I}.
$$

Integrating over $\varphi$ we get
$$
   I = 2 \pi \text{J}_0\left( 2\pi g_{\vec{k}} r \right).
$$
Then we substitute last equality and expression for $V(r)$ from \eqref{deformPotential} into equation for $V_{\vec{k}}$.
It is known~\cite{gradstein_ryzhyk}, if $a>0$ and $b>0$
\begin{equation}
 \int\limits_0^\infty x\text{J}_0(ax)\text{K}_0(bx)dx = \frac{1}{a^2+b^2}.
\end{equation}
With this equation we easily obtain $V_{n,m}$ for capillary interaction part
\begin{equation}
 V_{\vec{k}} = - \frac{q^2E^2}{\sigma} \frac{\bar{\rho}}{4\pi^2 g_{\vec{k}}^2 + 1/l_0^2}.
\end{equation}
Comparison with \eqref{capillary_v_k} shows that they are equal.

\subsection{Approximation of \texorpdfstring{$F_{int}$}{Fint} for capillary interaction}
\label{subsec:approximation_f_cap}

We start with expression~\eqref{f_int_cap} for $F_{int}$.
First of all we rewrite it as follows
$$
\begin{aligned}
 F_{int} &=-\frac{\bar{\rho} q^2 E^2}{2\sigma} e^{s^2/l_0^2}\!\!
            \int\limits_{s}^{\infty}\!\!
                \mathfrak{s}
                e^{-\mathfrak{s}^2/l_0^2}
                \sum_{\vec{k}\in\mathbb{Z}^2} e^{-4 \pi^2 \mathfrak{s}^2 \vec{k}^{\,T} \hat{\mathcal{G}} \hat{\mathcal{G}}^{T} \vec{k}}
            d\mathfrak{s}.
\end{aligned}
$$

Then we change summation to integration $\sum_{\vec{k}\in\mathbb{Z}^2} \rightarrow \iint_{\mathbb{R}^2}$.
There is a method of integration for this kind of functions \cite{bellman_matrix}.
As result we get
$$
 F_{int} =-\frac{\bar{\rho} q^2 E^2}{2\sigma} e^{s^2/l_0^2}
            \int\limits_{s}^{\infty}
                \mathfrak{s}
                e^{-\mathfrak{s}^2/l_0^2}
                \frac{d\mathfrak{s}}{2\mathfrak{s}^2 \det\hat{\mathcal{G}}}.
$$

For two-dimension density it is true $\det\hat{\mathcal{G}} = \bar{\rho}$.
Besides, we perform integration over $\mathfrak{s}$ and get
$$
 F_{int} =-\frac{q^2 E^2}{8\pi\sigma} e^{s^2/l_0^2}
           \Gamma_0\left(s^2/l_0^2\right),
$$
where $\Gamma_0$ is gamma-function.
Since w know that $s \ll l_0$ last equation can be approximated.
We use known approximation \cite{abramowitz_stegun} $\Gamma_0(x \to 0) \approx -\gamma -\ln(x)$, where $\gamma$ is Euler–Mascheroni constant.
We neglect this constant and exponent.
The rest can be written as \eqref{f_int_cap_final}.

\end{document}